\documentclass[aps,prd,preprint,showpacs,amsmath,amssymb,floatfix]{revtex4} 
\usepackage{epsfig} 
 
\newlength{\www}

\newcommand{\be}{\begin{equation}} 
\newcommand{\ee}{\end{equation}} 
\newcommand{\ba}{\begin{eqnarray}} 
\newcommand{\ea}{\end{eqnarray}} 
 
\newcommand{\bq}{\begin{equation}} 
\newcommand{\eq}{\end{equation}} 
\newcommand{\bqa}{\begin{eqnarray}} 
\newcommand{\eqa}{\end{eqnarray}} 
\newcommand{\ben}{\begin{enumerate}} 
\newcommand{\een}{\end{enumerate}} 
\newcommand{\bc}{\begin{center}} 
\newcommand{\ec}{\end{center}} 
\newcommand{\bqb}{\begin{eqnarray*}} 
\newcommand{\eqb}{\end{eqnarray*}}

\begin{document} 
 
\preprint{PTA/08-004} 
 
\title{\vspace{1cm} 
A complete one-loop calculation  of electroweak supersymmetric   
effects in  $t$-channel single top production at LHC 
}


\author{M.~Beccaria$^{a,b}$, C.~M.~Carloni Calame$^{c}$, 
G.~Macorini$^{d, e}$, E.~Mirabella$^f$,  
F.~Piccinini$^{g}$,  
F.~M.~Renard$^h$ and C.~Verzegnassi$^{d, e}$ \\ 
\vspace{0.4cm} 
} 
 
\affiliation{\small 
$^a$ Dipartimento di Fisica, Universit\`a del Salento \\ 
Via Arnesano, I-73100 Lecce, Italy\\ 
\vspace{0.2cm} 
$^b$ INFN, Sezione di Lecce, Italy\\ 
\vspace{0.2cm} 
$^c$ 
INFN, Italy and  School of Physics \& Astronomy, \\
University of Southampton, Southampton S017 1BJ, England\\ 
\vspace{0.2cm} 
$^d$ 
Dipartimento di Fisica Teorica, Universit\`a di Trieste, \\ 
Strada Costiera 
 14, Miramare (Trieste), Italy \\ 
\vspace{0.2cm} 
$^e$ INFN, Sezione di Trieste\\ 
$^f$ Max-Planck-Institut f\"ur Physik \\ F\"ohringer  
Ring 6 D-80805 M\"unchen, Germany \\  
\vspace{0.2cm} 
$^g$ INFN, Sezione di Pavia, \\
Via A. Bassi 6, I-27100 Pavia, Italy\\ 
\vspace{0.2cm}
$^h$ Laboratoire de Physique Th\'{e}orique et Astroparticules, 
UMR 5207\\ 
Universit\'{e} Montpellier II, 
 F-34095 Montpellier Cedex 5.\hspace{2.2cm}\\ 
\vspace{0.2cm} 
}


\begin{abstract} 
We have computed the complete one-loop electroweak effects in the MSSM  
for single top (and single antitop) production  
in the $t$-channel at hadron colliders, generalizing a previous analysis  
performed for the dominant $dt$ final state and fully including QED effects.  
The results are quite similar for all processes. The overall Standard Model  
one-loop effect is small, of the few percent size. This is due to a 
compensation of weak and QED contributions that are of opposite sign.  
The genuine SUSY contribution is generally quite modest in the mSUGRA 
scenario. The experimental observables would therefore only  
practically depend, in this framework, on the CKM $Wtb$ coupling. 
\end{abstract} 
 
 
\maketitle 
 
\section{Introduction} 
\label{sec:intro} 
 
%
%
%
%
 
The relevance of a precise measurement of single top production at hadron colliders  
was already stressed  in several papers in the recent years 
~\cite{Tait:1999cf}. 
A well known peculiarity of the process is actually the fact that it offers  
the unique possibility of a direct measurement of the $Wtb$ coupling $V_{tb}$, 
thus allowing severe tests of the conventionally assumed properties  
of the CKM matrix in the Minimal Standard Model; for a very accurate 
review of the topics we defer to~\cite{Alwall:2006bx}. 
 
For the specific purpose of a "precise" determination of $V_{tb}$,  
two independent requests must be met. The first one is that of a 
correspondingly "precise" experimental  measurement of the process. For 
the $t$-channel case on which we shall concentrate in this paper the CMS 
study~\cite{Pumplin:2002vw}  concludes that, with $10$ fb$^{-1}$ of  
integrated luminosity, one could be able to reduce the (mostly systematic)  
experimental uncertainty of the cross section below the ten percent 
level (worse uncertainties are expected for the two other processes, the 
$s$-channel and the associated $Wt$ production, whose cross section is 
definitely smaller than that of the $t$-channel).\\  
The second request is that of a similarly  
accurate theoretical prediction of the observables of the process.  
In this respect, one must make the precise statement that, in order to cope with the 
goal of measuring $V_{tb}$ at the few (five) percent level, a complete NLO  
calculation is requested. In the SM this has been done for the QCD component  
of the $t$-channel, resulting in a relatively small (few percent)  
effect~\cite{Stelzer:1997ns}.  
The electroweak effects have been computed very recently at the 
complete one loop level within the MSSM for the dominant $ub\to dt$ component of the process 
(to be defined in Section~\ref{sec:section2})~\cite{Beccaria:2006ir}.\\ 
One conclusion was that the genuine SUSY effect, for a set on mSUGRA benchmark 
points, was systematically modest, roughly at the one-two percent level.  
The SM contribution was computed in a preliminary way, i.e. only including  
the soft photon radiation to achieve cancellation of infrared singularities,  
ignoring the potentially relevant hard photon contribution.  
Actually, the main purpose of~\cite{Beccaria:2006ir} was that of investigating the possible  
existence and size of genuine supersymmetric effects, that would be 
essentially unaffected  by the SM QED contribution.  
In this approximate approach, the one-loop SM contribution turned out to be 
sizable, of roughly ten percent on the total rate. This contribution should be  
considered as a perfectly known term, to be included in the theoretical  
expression of the rate and compared with the corresponding experimental  
measurement to extract the precise value of the $Wtb$ coupling $V_{tb}$.\\ 
In fact, from the negative (for what concerns supersymmetric searches) result  
that the $t$-channel rate is not sensitive to genuine mSUGRA MSSM effects,  
adding the extra known feature that NLO QCD effects are small , of the order 
of five percent, and well under 
control~\cite{QCDNLO}~\cite{QCDNLODecay}~\cite{Frixione:2005vw},  
one concludes that an extremely precise theoretical determination of  
the process would be possible, provided that a rigorous electroweak  
one-loop description were given. This requires two different steps:  
first, an extension of the calculation of ~\cite{Beccaria:2006ir}  
to the seven remaining $t$-channel processes since the final $dt$ state is only the numerically dominant one. 
Second, the additional calculation of the complete QED effect,  
including properly hard photon radiation. 
This is precisely the aim of this paper, whose goal will be that of offering  
a clean theoretical expression to be used for a significant measurement of 
$V_{tb}$.\\ 
Technically speaking, this paper is organized as follows. In Section 
\ref{sec:section2} the definition of the eight considered  
processes will be given, and the main tasks that were fulfilled  
to perform a complete one loop electroweak calculation will be indicated.  
Since the problems to be solved were practically identical with  
those already met in~\cite{Beccaria:2006ir}, our description will  
be whenever possible quick and essential. 
In Section \ref{sec:QED}, we shall expose the new 
calculation of the complete QED 
effects.  
In Section \ref{sec:SUSYQCD} we briefly discuss the one loop SUSY QCD effects that we have recomputed 
from scratch in the same scheme adopted for the electroweak corrections. 
In Section \ref{sec:observable} we shall define the considered observables 
and show the results  
of our calculation, particular emphasis being given to the value of the  
total $t$-channel rate. 
A few conclusions will be drawn in the final Section \ref{sec:concl}.

\section{The $t$-channel processes at one electroweak loop} 
\label{sec:section2} 
The complete description of the $t$-channel  involves at 
partonic level four  sub-processes for single $t$ production:  $ub\to td$,  
$\bar{d}b\to t\bar{u}$,  $cb\to ts$,  $\bar{s}b\to t \bar{c}$,  
and the related four for the single $\bar{t}$ production.\\ 
The starting point is the cross section for the $ub\to td$ process 
with the complete set of one loop electroweak corrections (in the MSSM and SM).  
The $\mathcal{O}(\alpha)$ corrections to the (unpolarized) differential cross section of this process read: 
\begin{equation} 
d \sigma^{\mbox{\tiny ew}}_{ub\to td} =  \frac{dt}{64 \pi s^2} \sum_{\mbox{\tiny spin}}    
2  \mathfrak{Re} \{ \mathcal{M}^{0 ~*} \mathcal{M}^1  \} 
\end{equation} 
where $\mathcal{M}^0$ is the tree level contribution to the amplitude of the partonic process $ub \to dt$ while
$\mathcal{M}^1$ describes the EW one loop contribution to the same amplitude. The Mandelstam variables are defined as: 
\begin{eqnarray} 
s&=&(p_b+p_u)^2\nonumber\\ 
t &=& (p_b-p_t)^2 \\ 
u &=& (p_b-p_d)^2\nonumber  
\end{eqnarray} 
The analytical  expression for $\mathcal{M}^{0}$ is available in  
literature (see for instance~\cite{Beccaria:2006ir}). 
 $\mathcal{M}^{1}$  has been generated with the help of \verb|FeynArts|~\cite{FeynArts}, the algebraic reduction of the  
one loop integrals is performed with the help of \verb|FormCalc|~\cite{FormCalc} and 
the scalar one loop integrals are  numerically evaluated using \verb|LoopTools|~\cite{LoopTools}. 
We treat UV divergences using dimensional reduction while IR 
singularities are parametrized giving a small mass $m_\gamma$ to the photon. The masses of the light quarks  
are used as regulators of the collinear singularities and are set to zero elsewhere.  \\  
UV divergences are cured renormalizing the parameters and the wavefunctions appearing in $\mathcal{M}^{0}$. In our case we have to renormalize the wavefunction of the  
external quarks, the mass of the W boson, the electroweak mixing  angle and the electric charge. We use the on shell scheme decribed in Ref.~\cite{DennerHab}.  
This scheme uses the fine structure constant evaluated in the Thomson limit as input parameter. In order to avoid large logarithms arising from the runnning  
of $\alpha$ to the electroweak scale $M_W$, we slightly modify this scheme using as input parameter the Fermi constant $G_F$. We consistently change the definition 
of the renormalization constant of the fine structure constant following the guidelines of Ref.~\cite{WjetProd}. \\ 
 
\noindent 
The unpolarized differential cross section for the process  $\bar d b \to \bar u t$ can be obtained from that of the process $ub\to td$ by crossing: 
\begin{equation} 
d \sigma^{\mbox{\tiny ew}}_{\bar d b \to \bar u t} = \frac{dt}{64 \pi s^2} \sum_{\mbox{\tiny spin}}    
2  \mathfrak{Re} \{ \mathcal{M}^{0 ~*}(s\to u,~u\to s)  \mathcal{M}^1(s\to u,~u\to s)  \}   . 
\end{equation} 
For the $\bar{t}$ production the cross sections can be calculated using the 
identities 
\bqa 
d \sigma^{\mbox{\tiny ew}}_{u b \to t d} &=& d \sigma^{\mbox{\tiny ew}}_{\bar{u}  \bar{b} \to \bar{d} \bar{t}}\nonumber\\ 
d \sigma^{\mbox{\tiny ew}}_{\bar{d} b \to t \bar{u}} &=& d \sigma^{\mbox{\tiny ew}}_{d  \bar{b} \to u \bar{t}} 
\eqa 
while the  processes involving the second generation $c,s$ and $\bar{s}, \bar{c}$ 
quarks can be computed from the previous, simply replacing the masses of the external particles 
(and some masses in the loop corrections).

\section{QED effects} 
\label{sec:QED} 
\begin{figure}[htb] 
\vspace{2.0cm} 
\centering 
\epsfig{file=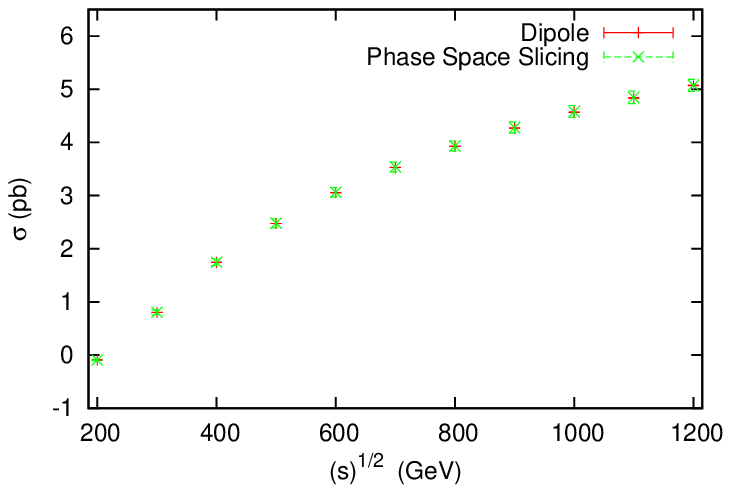, width= 7.9cm, angle=0} 
\epsfig{file=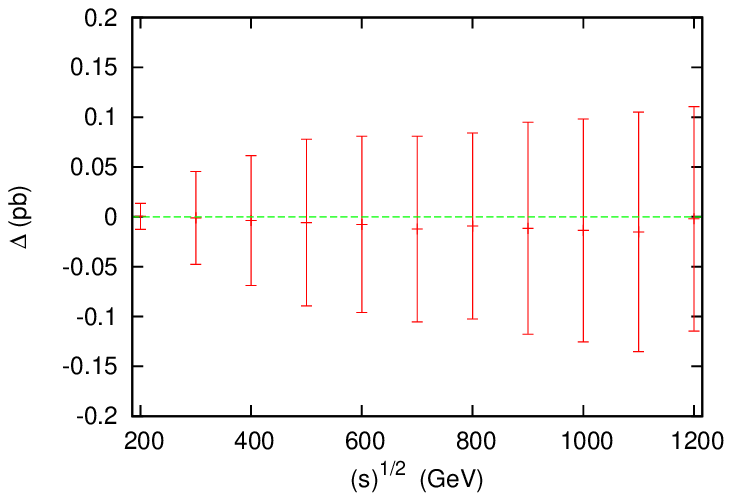,   width= 7.9cm, angle=0} 
\caption{Lowest order partonic cross section for the process $u b\to d t \gamma$ computed with the two different methods. 
The cuts for the Phase Space Slicing methos are $\delta_s = \delta_c = 10^{-4}$. The quantity $\Delta$ is defined  
as  $\Delta= \sigma^{\mbox{\tiny Dipole}} - \sigma^{\mbox{\tiny Slicing}}$.} 
\label{fig:2} 
\end{figure}

In order to obtain physically meanigful observables one has to include the differential cross section for  
the process of $t$-channel single top production associated with the 
emission of a  photon integrated over the whole photonic phase space. 
So we have to consider the partonic processes 
$ub\to td \gamma$, $\bar{d}b\to t\bar{u} \gamma$,  $cb\to ts \gamma$,  $\bar{s}b\to t \bar{c} \gamma$, 
and the related four for the single $\bar{t}$ production.  \\ 
The unpolarized differential cross section of these processes has been obtained using two different procedures. In the first  
approach the amplitude has been generated and squared using \verb|FeynArts| and  \verb|FormCalc| while in the second one  
the complete matrix element has been calculated with the help of FORM. The two methods are in mutual agreement. \\ 
The phase space integration of the aforementioned differential cross section is singular in the region in which the photon is soft  
and in the region in which it is collinear to a massless quark. 
According to KLN theorem~\cite{KLN} IR singularities and the 
collinear singularities  
related to the final state radiation cancel in sufficiently inclusive 
observables while the collinear singularities related to initial state 
radiation 
have to be absorbed in the redefinition of the Parton Distribution 
Functions (PDF). In order to 
regularize these divergences   
we use two different procedures: the Dipole Subtraction method and 
the Phase Space Slicing method. \\ 
In the Subtraction approach one has to  add and subtract 
to the squared amplitude  
an auxiliary function with the  same asymptotic behaviour and 
such that it can be analytically integrated 
over the photon phase space. Among the different choices we 
use the function quoted in Ref.~\cite{Dipole}. 
In this Reference   
explicit expression for the subtraction function and for its  
analytical integration is obtained using  
the so called Dipole Formalism~\cite{DipoleQCD}. \\ 
The idea behind the Phase Space Slicing approach is 
to isolate the singular region of the phase 
space introducing a cut off on the energy of the 
photon ($\Delta E = \delta_s \sqrt{s} /2$) and on the 
angle between the photon and  
the massless quarks ($\delta_c$). In the regular region the phase space integration can be done numerically while in the singular region  
it can be performed analytically provided that the cutoffs are small enough. The form of the differential cross section in the  
singular region is universal and its explicit expression in the soft (collinear) region  can be  
found in Ref.~\cite{DennerHab}~(\cite{WjetProd}). \\ 
As can be inferred from Fig.~\ref{fig:2} the two methods are in good  numerical agreement.

\section{One loop SUSY QCD corrections} 
\label{sec:SUSYQCD} 
 
The one loop SUSY QCD corrections to $t$-channel single top production have been computed at LHC in Ref.~\cite{Zhang:2006cx}. 
We include these corrections re-computing them from scratch following the guidelines described in Sec.~\ref{sec:section2}. 
The only difference is that, following a standard procedure in SUSY QCD, we treat UV divergences using dimensional regularization. 
Moreover in  this case we have to renormalize 
only the wavefunctions of the squarks since the other renormalization constant do not  
have $\mathcal{O}(\alpha_s)$ corrections. These corrections are IR safe.

\section{Observable quantities} 
\label{sec:observable} 
  
The differential hadronic cross section reads 
\begin{eqnarray} 
d \sigma (S) = \sum_{\mbox{\tiny{(q, q')}}}  \int_{\tau_0}^1 d \tau   
&& \frac{dL_{qb}}{d \tau}           \left(  
                                 d \sigma^{\mbox{\tiny ew}}_{q b \to q' t}(s)          +  
                                 d \sigma^{\mbox{\tiny ew}}_{q b \to q' t \gamma}(s)    + 
                                 d \sigma^{\mbox{\tiny SQCD}}_{ q b \to q' t }(s) \right)  \nonumber \\ 
&&  \frac{dL_{\bar q \bar b}}{d \tau} \left(  
                                 d \sigma^{\mbox{\tiny ew}}_{\bar q \bar b \to \bar q' \bar t}(s)          +  
                                 d \sigma^{\mbox{\tiny ew}}_{\bar q \bar b \to \bar q' \bar t \gamma}(s)    +  
                                 d \sigma^{\mbox{\tiny SQCD}}_{\bar q \bar b \to \bar q' \bar t }(s)     \right)  
\end{eqnarray} 
Where $(q,q')=(u,d),(c,s),(\bar d, \bar u), (\bar s, \bar d)$. The differential luminosity has been defined according to Ref.~\cite{PinkBook} 
while $\tau_0 = m^2_t / S$ and $s = \tau S$. $ d \sigma^{\mbox{\tiny ew}}_{X}$ ($d \sigma^{\mbox{\tiny SQCD}}_{X}$) 
are the $\mathcal{O}(\alpha^3)$ (SUSY QCD) corrections to  the differential cross section  of the process $X$. 
 
As pointed out in Sec.~\ref{sec:QED}, initial state Collinear Singularities  are not cancelled in the sum of virtual and real corrections and they are absorbed in the 
definition of the PDF. We use the $\overline{\mbox{MS}}$ factorization scheme at the scale $\mu_F = m_t$.  Concerning the choice of the parton distributions set, 
we follow ~\cite{WjetProd}. The calculation of the full $\mathcal{O}(\alpha)$ corrections to any hadronic observable must include  
QED effects in the DGLAP evolution equations. Such effects are taken into account in the MRST2004QED PDF~\cite{Martin:2004dh} which are NLO QCD.  
Since our computation is leading order QCD, we use the LO set CTEQ6L since the QED effects are known to be small~\cite{Roth:2004ti}. 
 
\subsection{Numerical Results} 
\label{subsec:num} 
 
In this subsection we present our numerical results. 
We define $M_{inv}$ as the invariant mass of the (anti)top quark and of the light
quark in  the final  state. Also, we denote by $p_T$ the transverse momentum of the  (anti)top quark.
We consider as physical observables the transverse momentum distribution 
$\frac{d \sigma}{dp_{T}}$, the invariant mass distribution $\frac{d 
 \sigma}{dM_{\mbox{\tiny inv}}}$  and two integrated observables derived from
the previous: the integrated transverse momentum distribution, defined as the 
integral of
$\frac{d \sigma}{dp_{T}}$ from a minimum $p_T^{\rm min}$ up to infinity:
\begin{equation}
\sigma(p_T^{\rm min}) = \int_{p_T^{\rm min}}^{\infty} ~\frac{d \sigma}{d p'_T} d p'_T
\end{equation}
and the cumulative invariant mass distribution 
$\sigma(M^{\rm max}_{\mbox{\tiny inv}})$, defined as the integral of the invariant mass distribution from 
the production threshold up to $M^{\rm max}_{\mbox{\tiny inv}}$: 
 
\begin{equation} 
\sigma(M^{\rm max}_{\mbox{\tiny inv}}) = 
\int_{\tau_0}^{M^{\rm max}_{\mbox{\tiny inv}}} ~\frac{d \sigma}{d M'_{\mbox{\tiny inv}}} d M'_{\mbox{\tiny inv}}
\end{equation} 
 
For each observable we present the plots for the LO and NLO curves and for the 
percentage effect of the NLO corrections to the observable $ (\delta = \frac{NLO 
  - LO}{NLO} \times 100)$. Apart from the Standard Model results, we  
analyzed several supersymmetric mSUGRA benchmark points: in the figures we
present the numerical results for the two representative 
ATLAS DC2 mSUGRA benchmark points SU1 and SU6~\cite{DC2}; 
in the other cases the results are similar.
The SU1 and SU6 input parameters generate a moderately light supersymmetric 
scenarios, where the masses of the supersymmetric particles are below the $1~TeV$:
the two physical spectra are quite similar, characterized by relatively
heavy squark sector ($562-631~GeV$ for the lightest squark in SU1 and SU6
respectively), a light neutralino and a light chargino state. The typical
masses for the supersymmetic Higgs are of order of $400-500~GeV$.
The main difference between the two points is the value of $\tan \beta$ 
input parameter: $10$ for SU1 and $50$ for SU6.   
\subsubsection{SM Results} 
 
\begin{figure}[htb] 
\epsfig{file=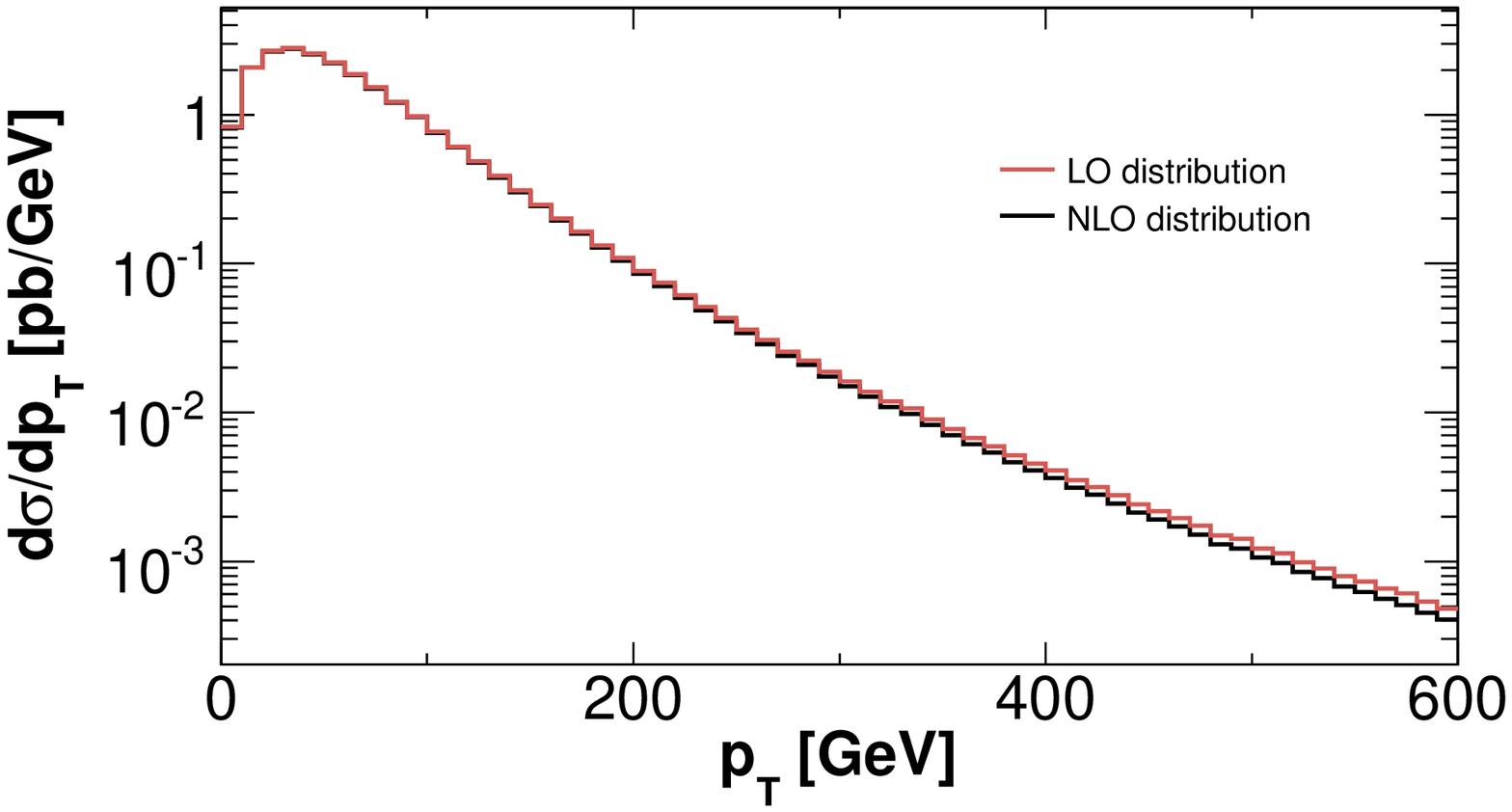,  width=8cm} 
\epsfig{file=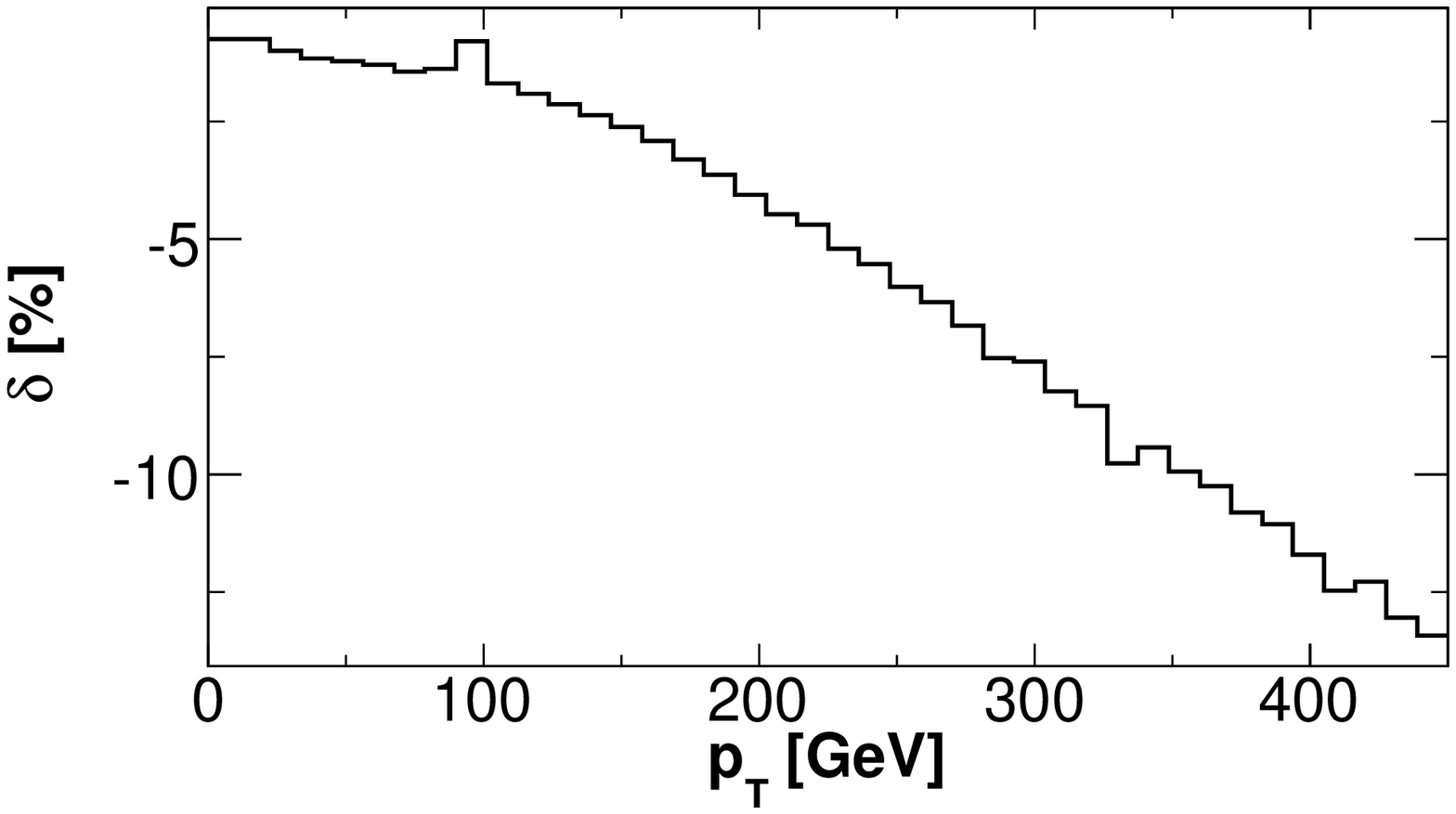, width=8cm} 
\caption{\footnotesize{
Left Panel: We plot the LO (that is tree level) 
contribution and the NLO (that is tree level plus  
$\mathcal{O}(\alpha^3)$) 
corrections to the transverse momentum  
distribution.\\
Right Panel: We plot the percentage contribution of  
the $\mathcal{O}(\alpha^3)$  corrections to the transverse momentum  
distribution; that is $\delta = \frac{NLO -LO}{NLO}\times 100$.\\  
No cuts are imposed. Computation in the Standard Model framework
}}  
\label{Fig:pT_SM} 
\end{figure}

\begin{figure}[htb] 
\vspace{-2cm} 
\centering 
\epsfig{file=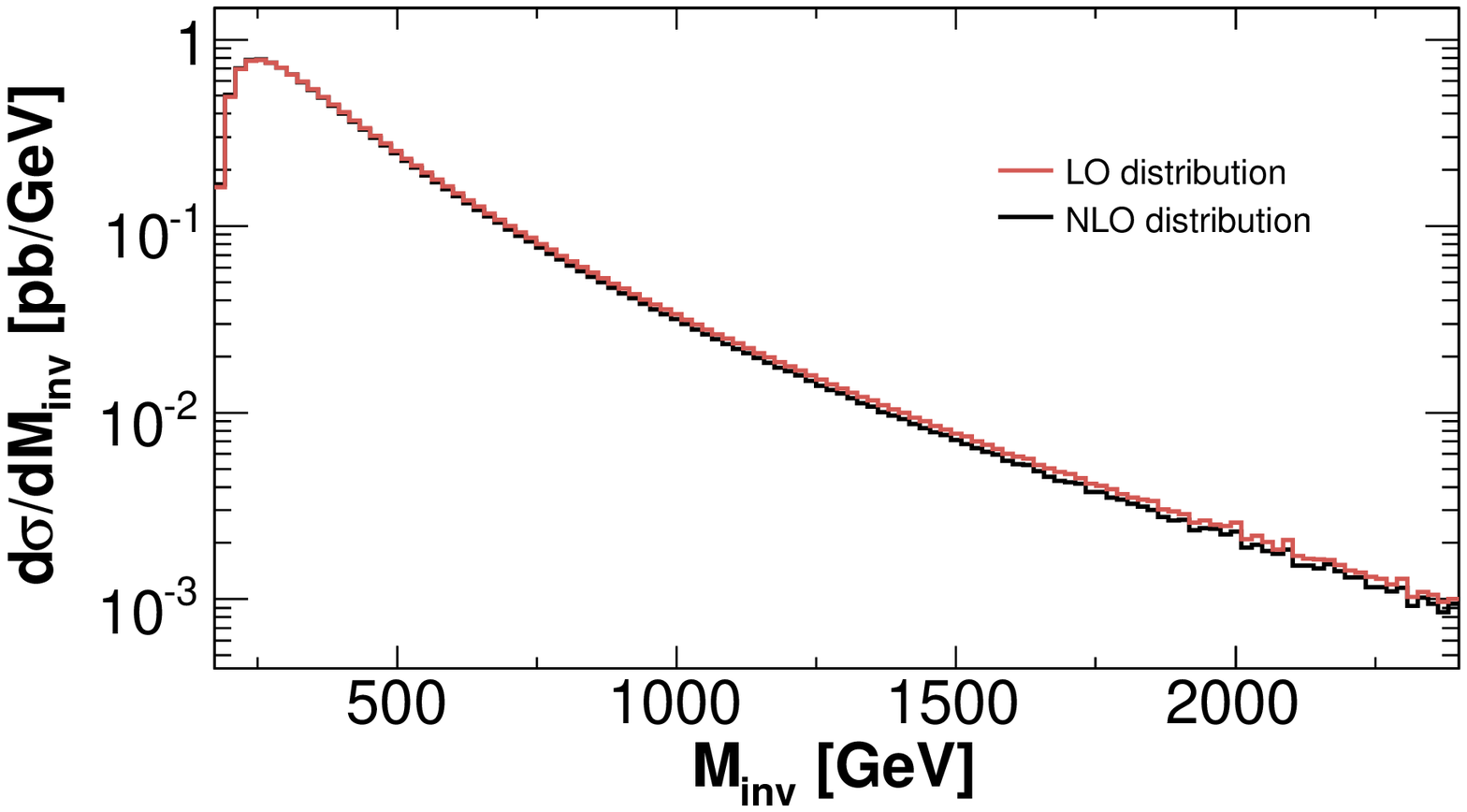 , width=8cm}
\epsfig{file=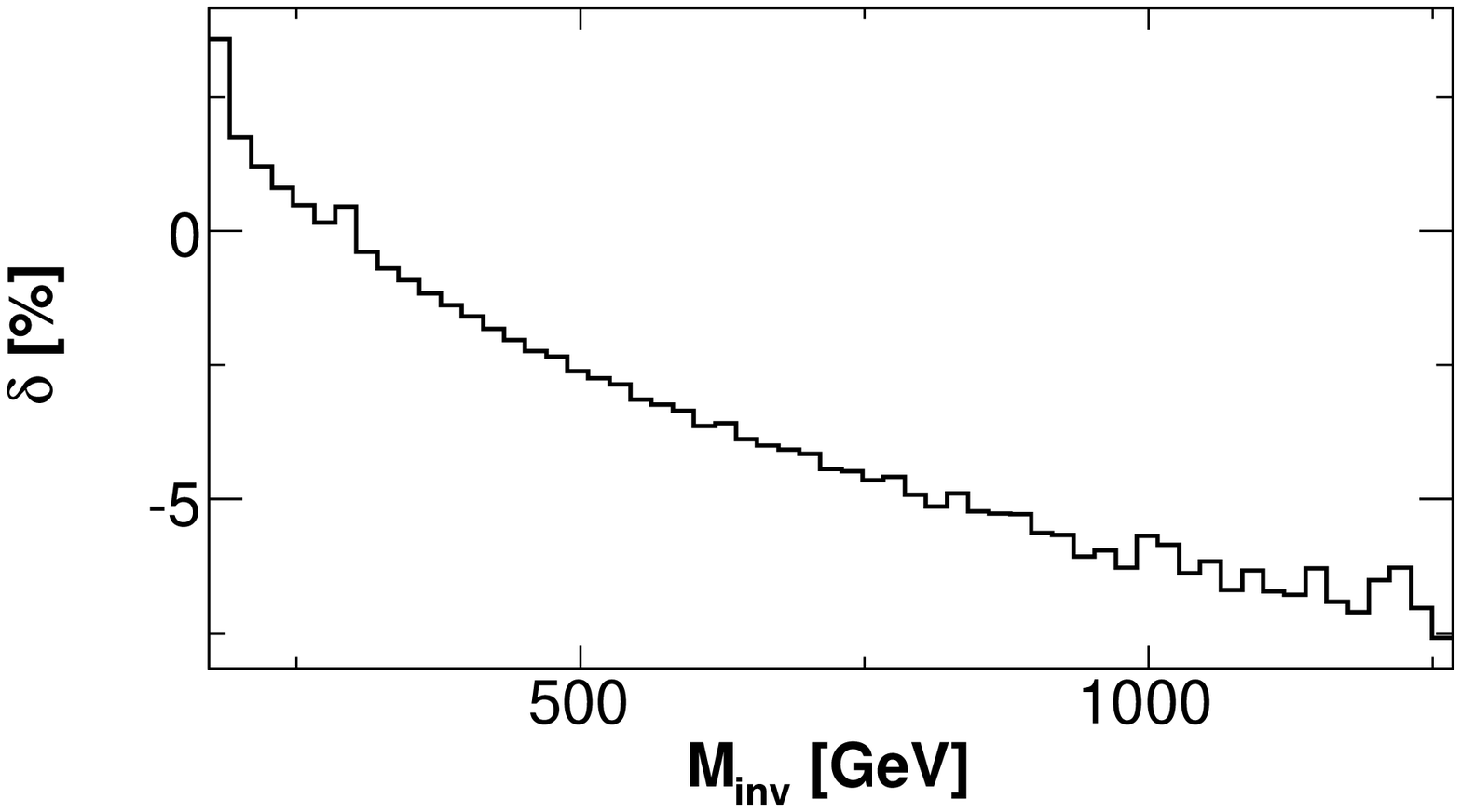, width=8cm} 
\caption{\footnotesize{
Left Panel: We plot the LO (that is tree level) 
contribution and the NLO (that is tree level plus $\mathcal{O}(\alpha^3)$) 
corrections to the invariant mass distribution.\\  
Right Panel: We plot the percentage contribution of  
the $\mathcal{O}(\alpha^3)$  corrections to the invariant mass   
distribution; that is $\delta = \frac{NLO -LO}{NLO}\times 100$.\\  
No cuts are imposed. Computation in the Standard Model framework}} 
\label{Fig:M_distr_SM} 
\end{figure}

\begin{figure}
\centering
\epsfig{file=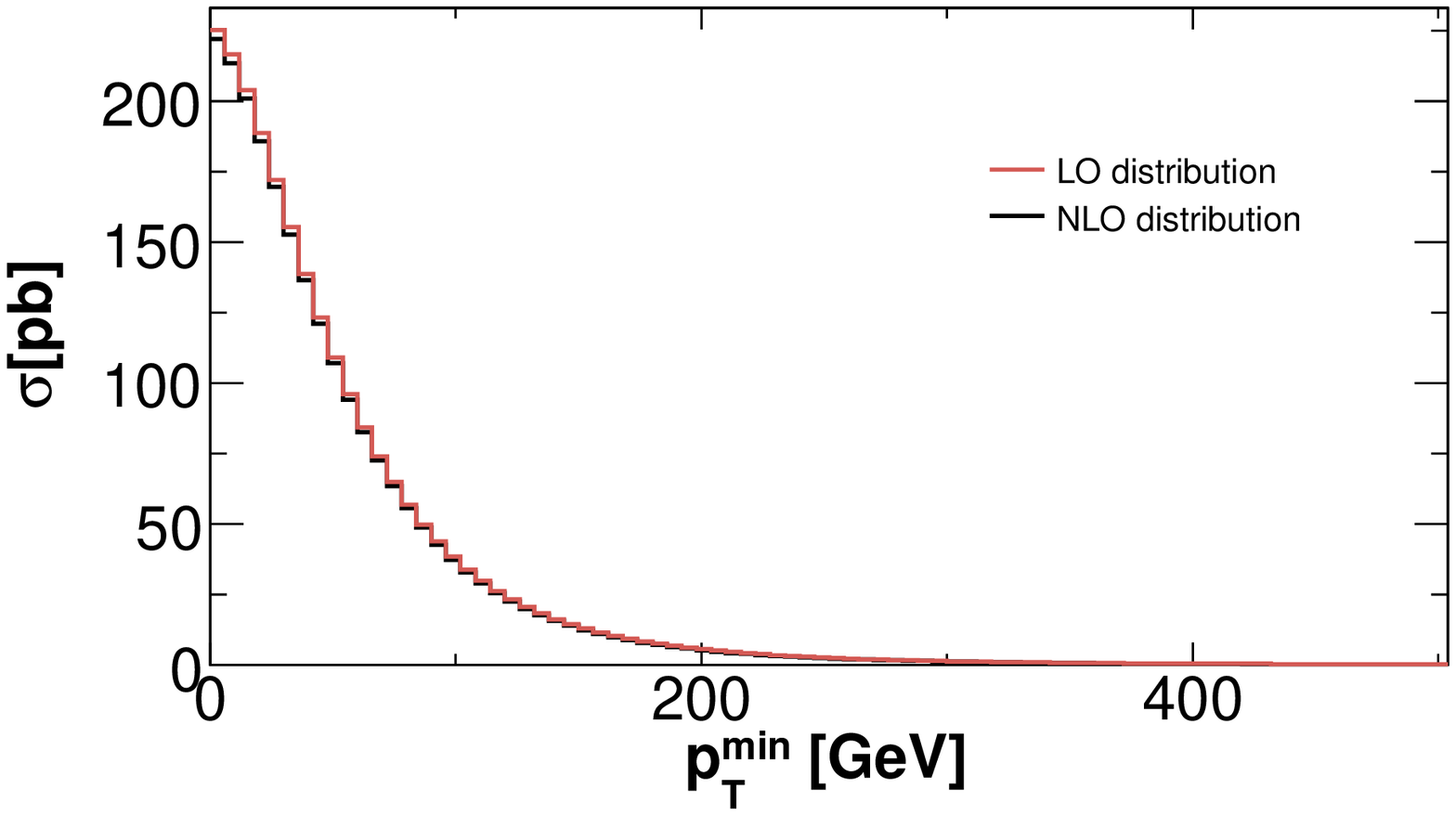 , width=8cm}
%
\epsfig{file=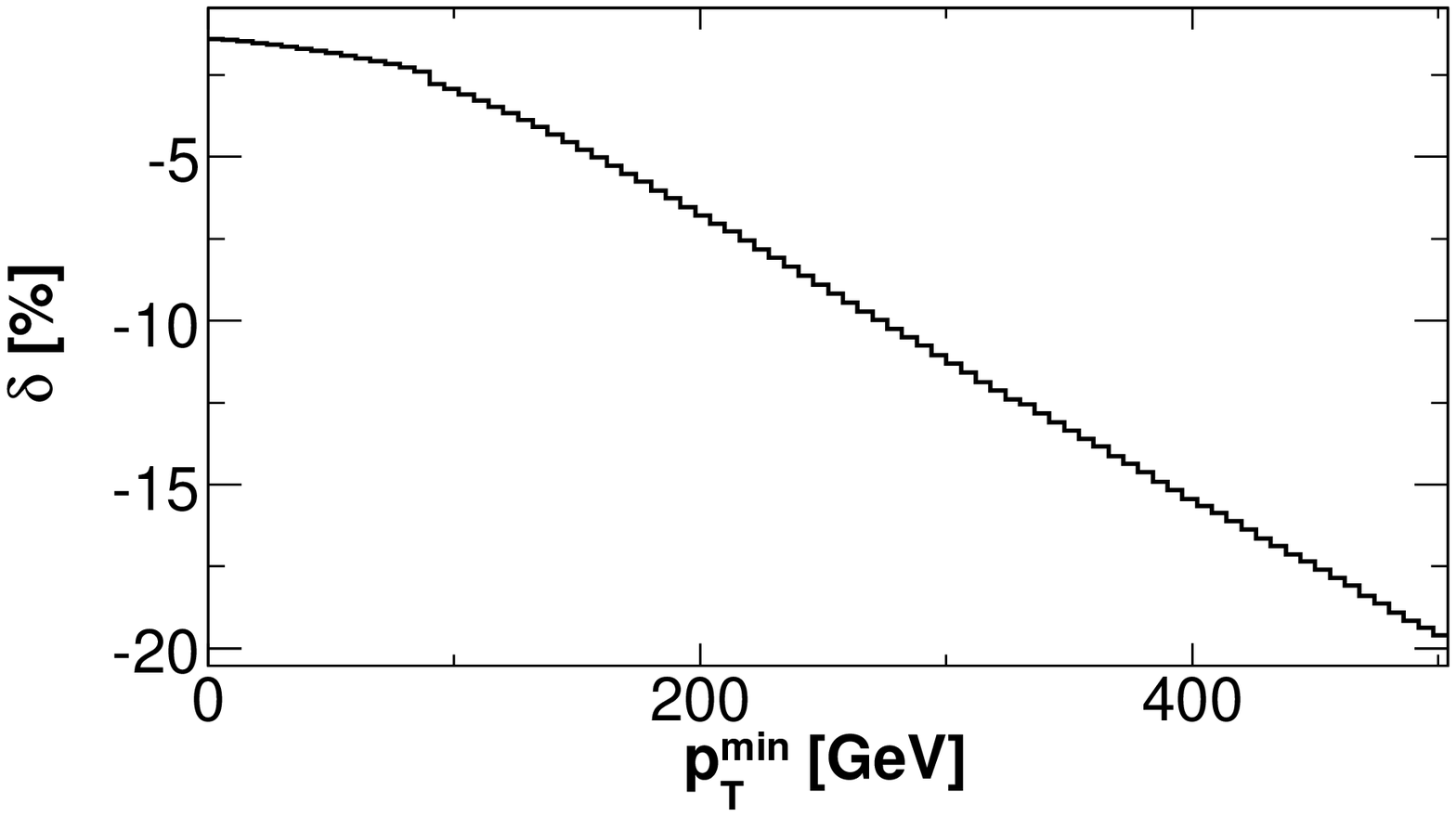, width=8cm}
\caption{ \footnotesize{
Left Panel: We plot the LO (that is tree level)
contribution and the NLO that is tree level plus 
$\mathcal{O}(\alpha^3)$
corrections to the integrated transverse momentum distribution $\sigma(p_T^{\rm min})$.
We remind that this distribution  is defined as the transverse momentum distribution  
integrated from a minimum $p_T^{\rm min}$ up to infinity.\\ 
%
Right Panel: We plot the percentage contribution of 
the $\mathcal{O}(\alpha^3)$  correections to the integrated transverse momentum 
distribution; that is $\delta = \frac{NLO -LO}{NLO}\times 100$. 
No cuts are imposed.Computation in the Standard Model framework}}
\label{Fig:pT_int_SM}
\end{figure}

\begin{figure}[htb] 
\centering 
\epsfig{file=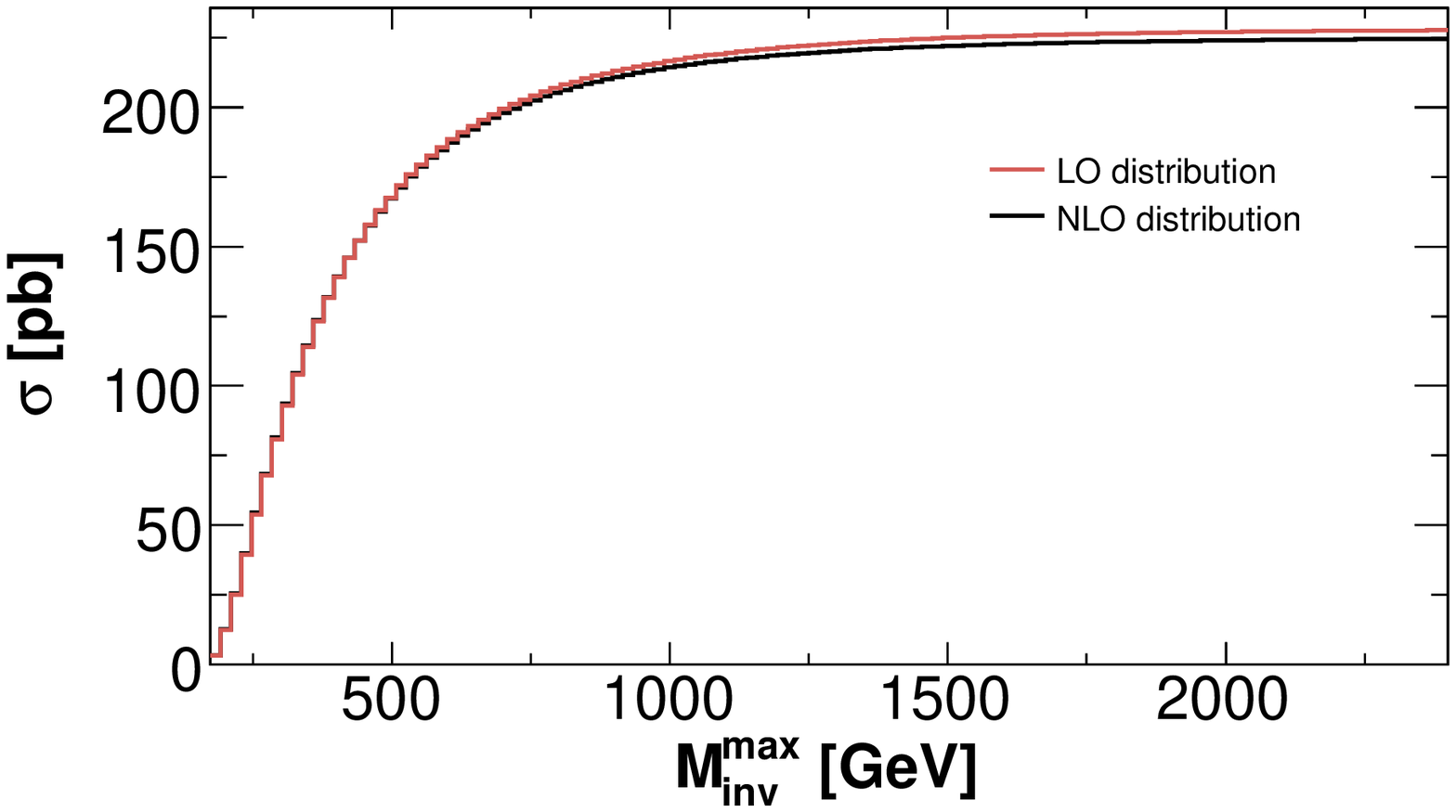 , width=8cm}
\epsfig{file=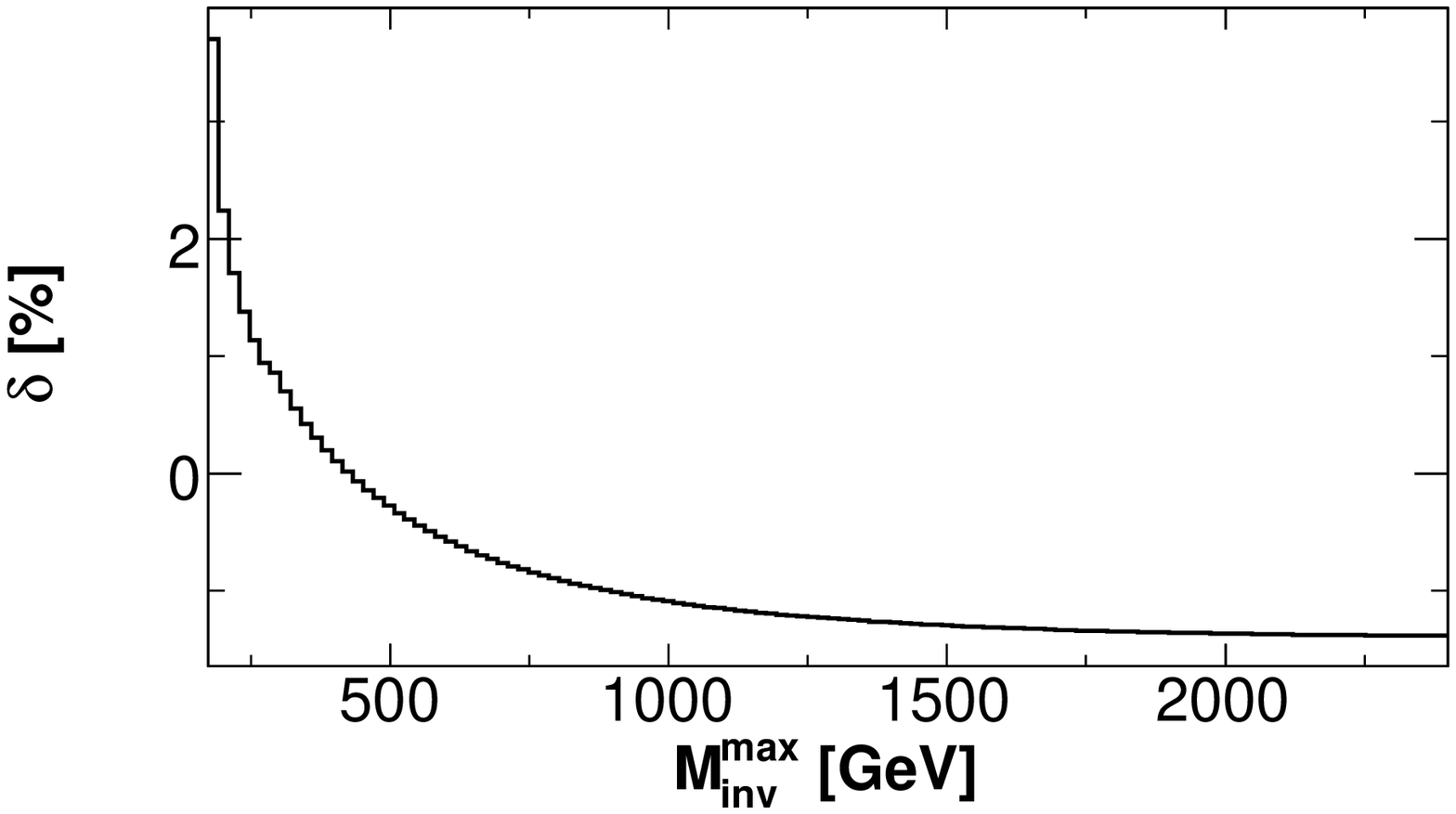, width=8cm}  
\caption{\footnotesize{
Left Panel: We plot the LO (that is tree level) 
contribution and the NLO (that is tree level plus $\mathcal{O}(\alpha^3)$) 
corrections to the cumulative invariant mass distribution $\sigma(M^{\rm max}_{\mbox{\tiny inv}})$. 
We remind that this distribution  is defined as the invariant mass distribution  
integrated from threshold up to $M^{\rm max}_{\mbox{\tiny inv}}$.\\ 
Right Panel: We plot the percentage contribution of  
the $\mathcal{O}(\alpha^3)$  corrections to the cumulative invariant mass   
distribution; that is $\delta = \frac{NLO -LO}{NLO}\times 100$.\\  
No cuts are imposed. Computation in the Standard Model framework}}  
\label{Fig:M_int_SM} 
\end{figure}

In the Standard Model framework the behaviour of the four observables is shown
in Figures \ref{Fig:pT_SM}, \ref{Fig:M_distr_SM}, \ref{Fig:pT_int_SM} and 
\ref{Fig:M_int_SM}: the left panels report the LO and NLO curves, and the right
panels show the relative percentage effect of the one loop electroweak
corrections: as one can see the curves for the NLO and LO are 
almost overlapping, and the global $\mathcal{O}(\alpha^3)$ NLO effect is rather 
small. This is particularly evident considering the plot for the cumulative 
invariant mass distribution, where the percentage effect saturate to the 
$\simeq 1.5 \% $ for $M_{\mbox{\tiny inv}} > 1500~GeV$.

\subsubsection{MSSM Results} 
 
\begin{figure}[htb] 
\centering 
{}
{\hspace{1cm} {\bf (a)} \hspace{8cm} {\bf (b)}}\\

\epsfig{file=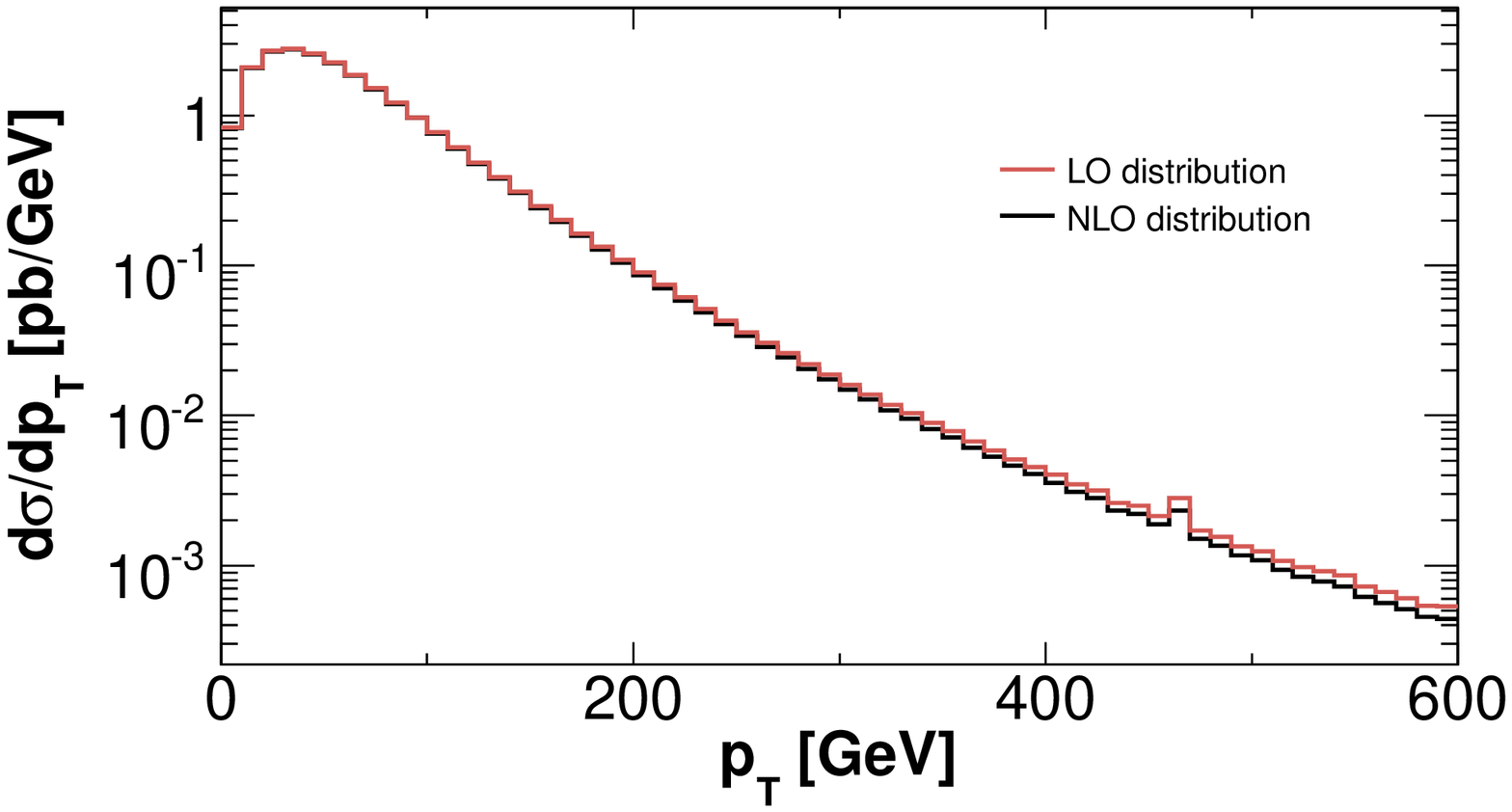 , width=8cm}
\epsfig{file=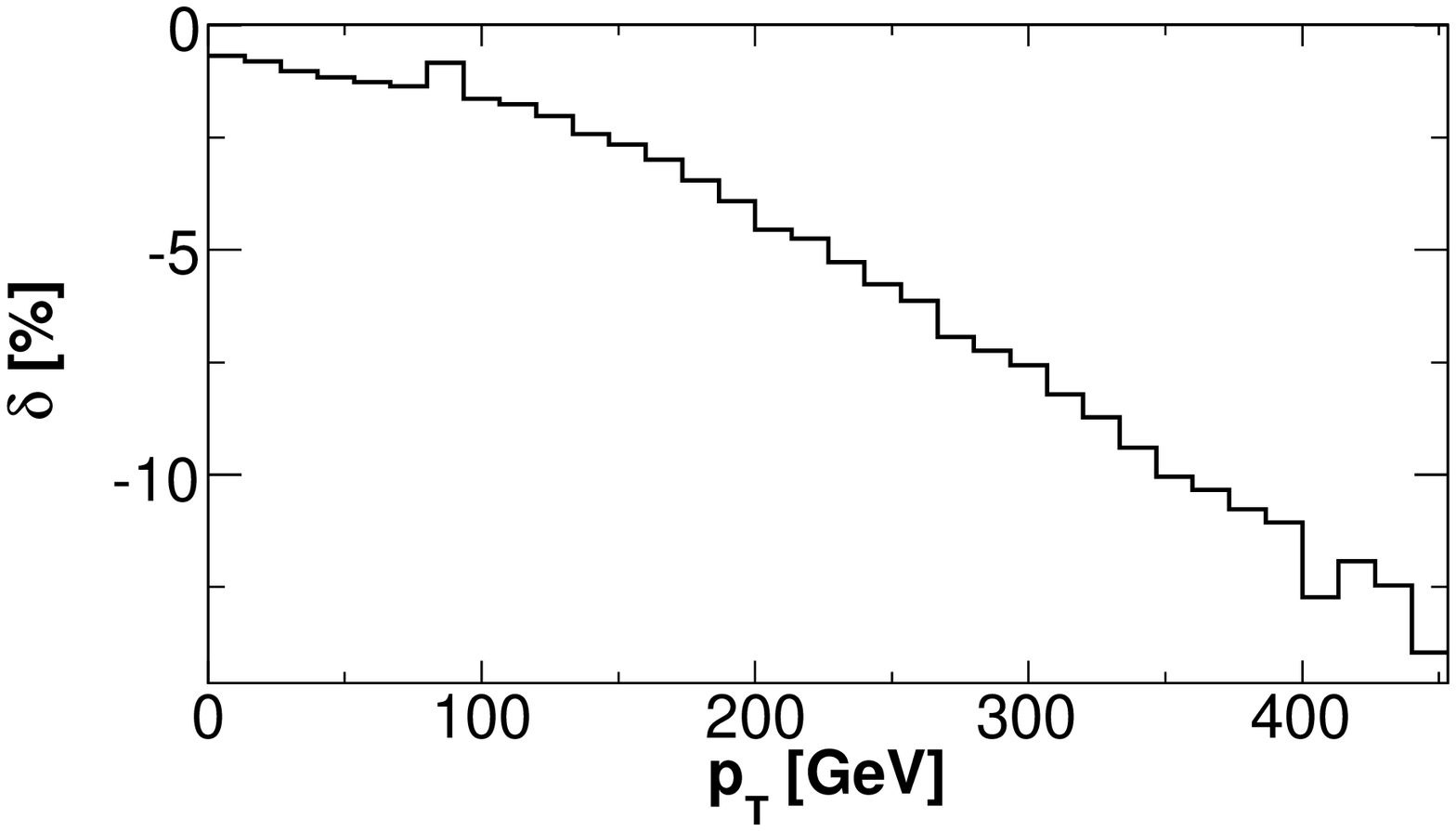, width=8cm}

{\hspace{1cm} {\bf (c)} \hspace{8cm} {\bf (d)}}\\

\epsfig{file=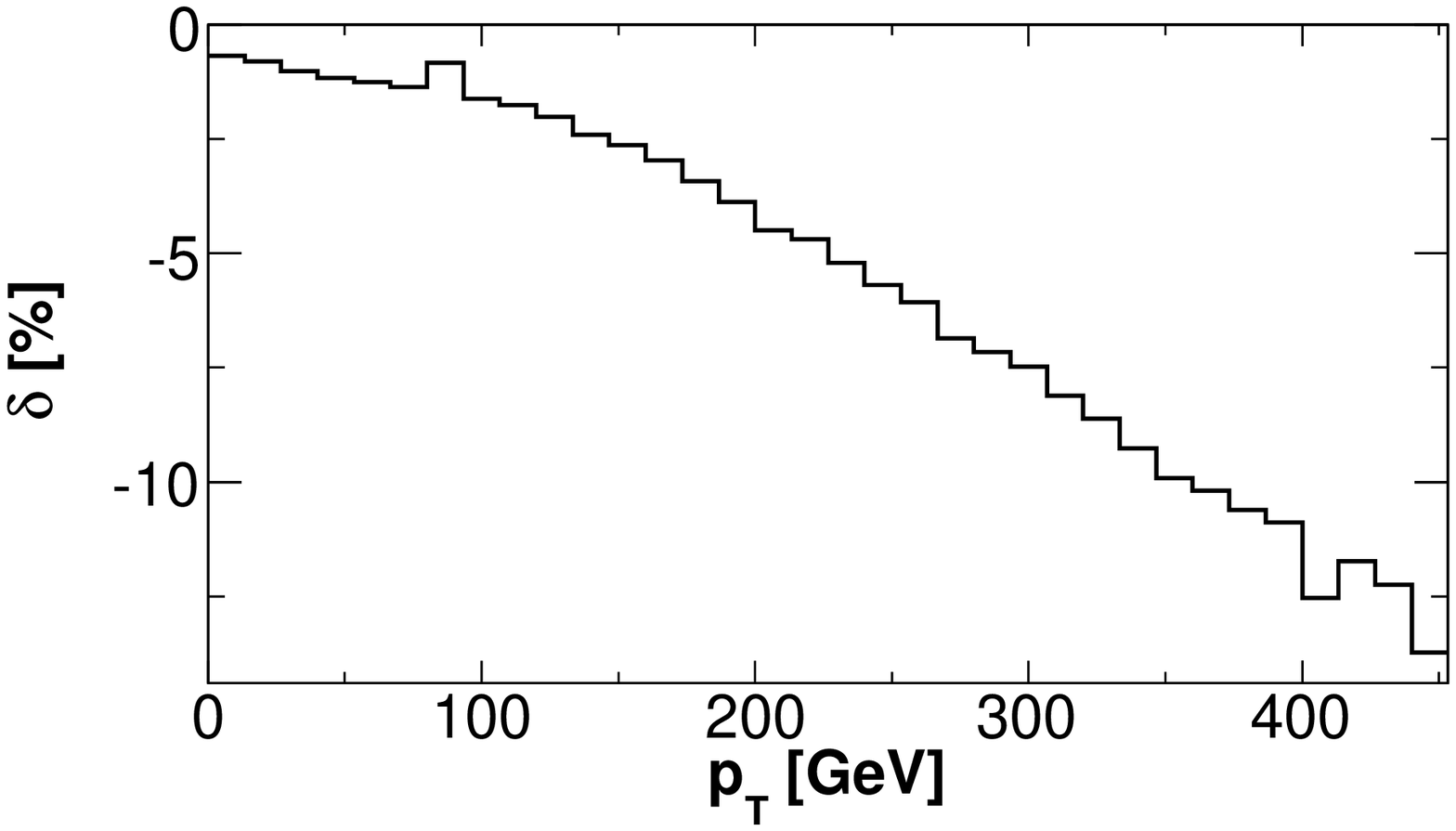, width=8cm} 
\epsfig{file=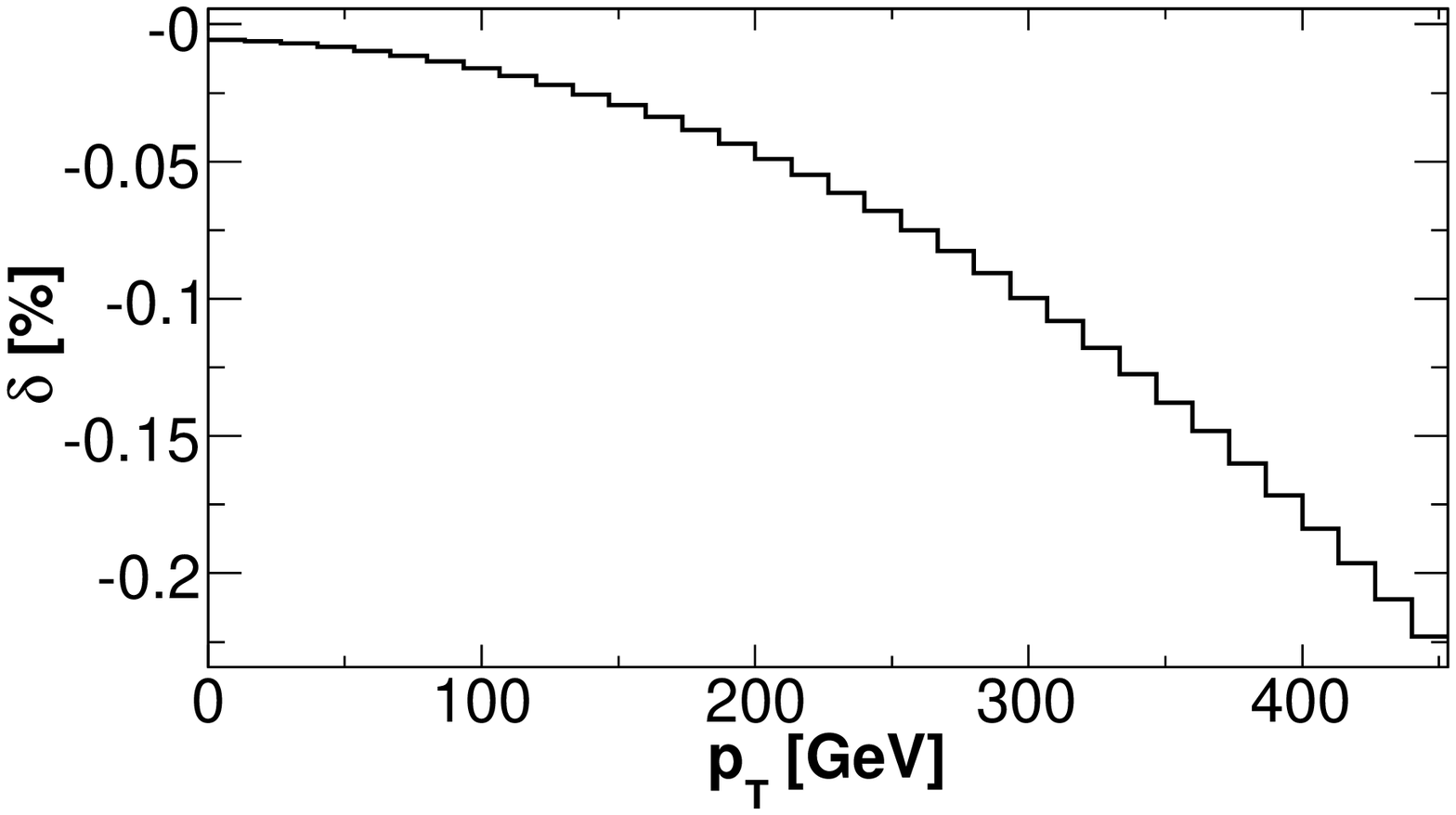, width=8cm}
\caption{\footnotesize{
{\bf(a) } We plot the LO (that is tree level) 
contribution and the NLO (that is tree level plus  
$\mathcal{O}(\alpha^3)$ plus SUSY QCD)   
corrections to the transverse momentum  
distribution.\\ 
%
{\bf (b)} We plot the percentage contribution of  
the $\mathcal{O}(\alpha^3)$ plus SUSY QCD   corrections to the transverse momentum  
distribution; that is $\delta = \frac{NLO -LO}{NLO}\times 100$.\\  
%
{\bf (c)} We plot the percentage contribution of  
the $\mathcal{O}(\alpha^3)$   corrections to the transverse momentum  
distribution; that is $\delta = \frac{\mathcal{O}(\alpha^3)}{NLO}\times 100$.\\  
%
{\bf (d)} We plot the percentage contribution of  
the SUSY QCD   corrections to the transverse momentum  
distribution; that is $\delta = \frac{SUSY~QCD}{NLO}\times 100$.\\  
${}$\\
No cuts are imposed. Computation in the SU1 point}
}  
\label{Fig:pT_SU1} 
\end{figure}

\begin{figure}[htb] 
\centering 
{}
{\hspace{1cm} {\bf (a)} \hspace{8cm} {\bf (b)}}\\

\epsfig{file=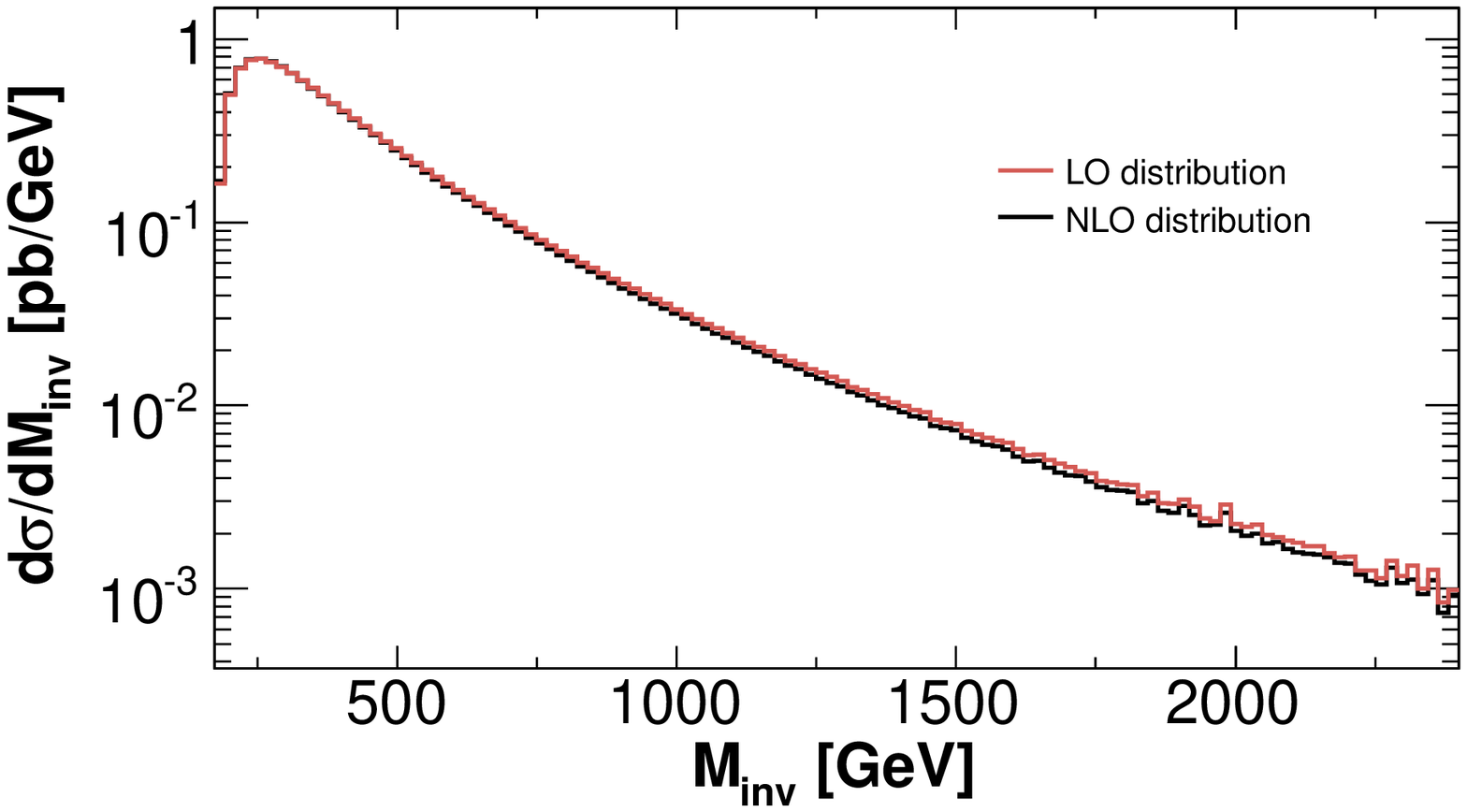 , width=8cm} 
\epsfig{file=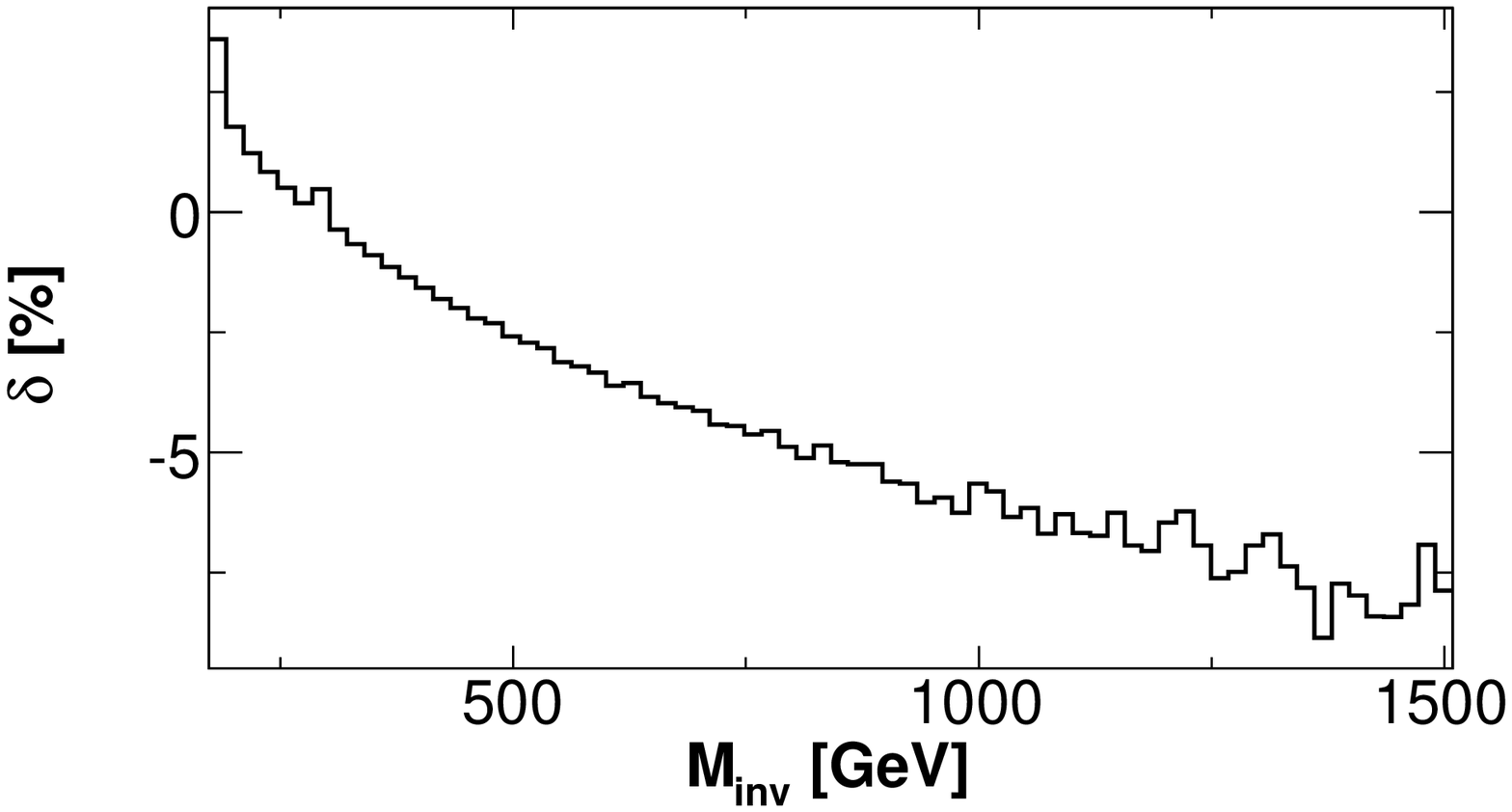, width=8cm} 

{\hspace{1cm} {\bf (c)} \hspace{8cm} {\bf (d)}}\\

\epsfig{file=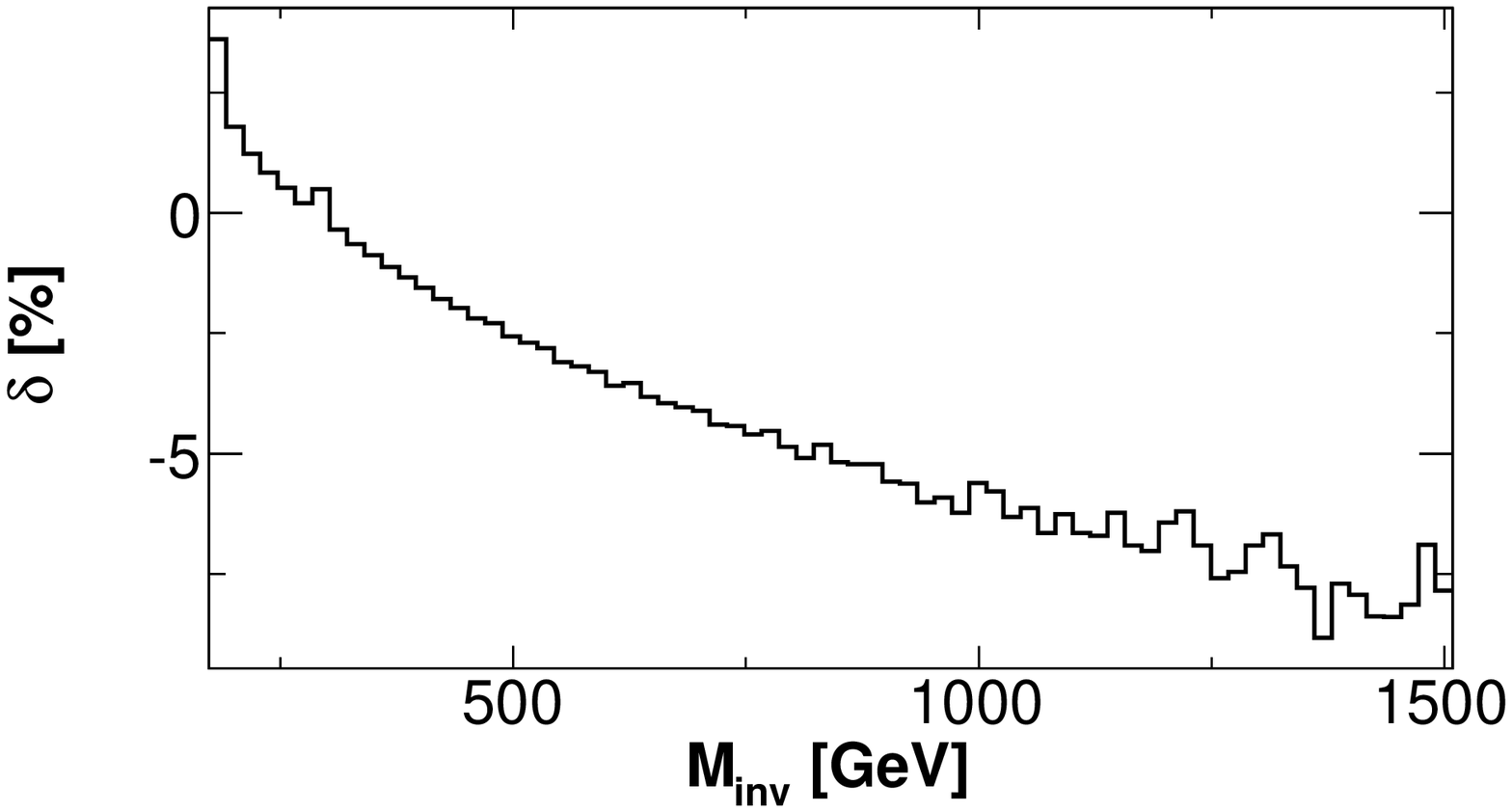, width=8cm} 
\epsfig{file=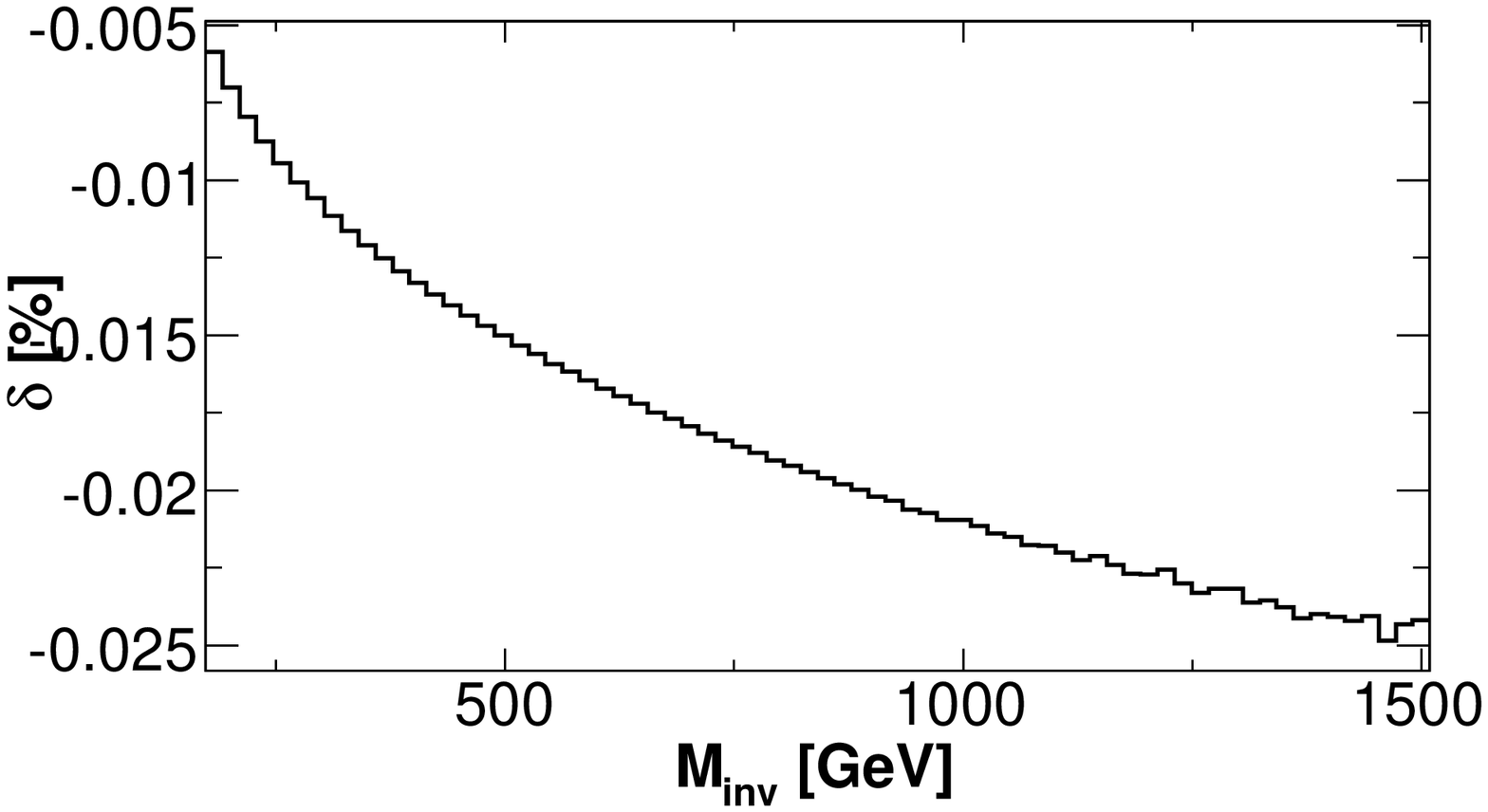, width=8cm} 
\caption{\footnotesize{
{\bf (a)} We plot the LO (that is tree level) 
contribution and the NLO (that is tree level plus $\mathcal{O}(\alpha^3)$ plus SUSY QCD) 
corrections to the invariant mass distribution.\\ 
%
{\bf (b)} We plot the percentage contribution of  
the $\mathcal{O}(\alpha^3)$ plus SUSY QCD   corrections to the invariant mass   
distribution; that is $\delta = \frac{NLO -LO}{NLO}\times 100$.\\  
%
{\bf (c)} We plot the percentage contribution of  
the $\mathcal{O}(\alpha^3)$   corrections to the invariant mass   
distribution; that is $\delta = \frac{\mathcal{O}(\alpha^3)}{NLO}\times 100$.\\  
%
{\bf (d)} We plot the percentage contribution of  
the SUSY QCD   corrections to the invariant mass   
distribution; that is $\delta = \frac{SUSY~QCD}{NLO}\times 100$.\\  
${}$\\
No cuts are imposed. Computation in the SU1 point
}} 
\label{Fig:M_distr_SU1} 
\end{figure}

\begin{figure}
\centering
{}
{\hspace{1cm} {\bf (a)} \hspace{8cm} {\bf (b)}}\\

\epsfig{file=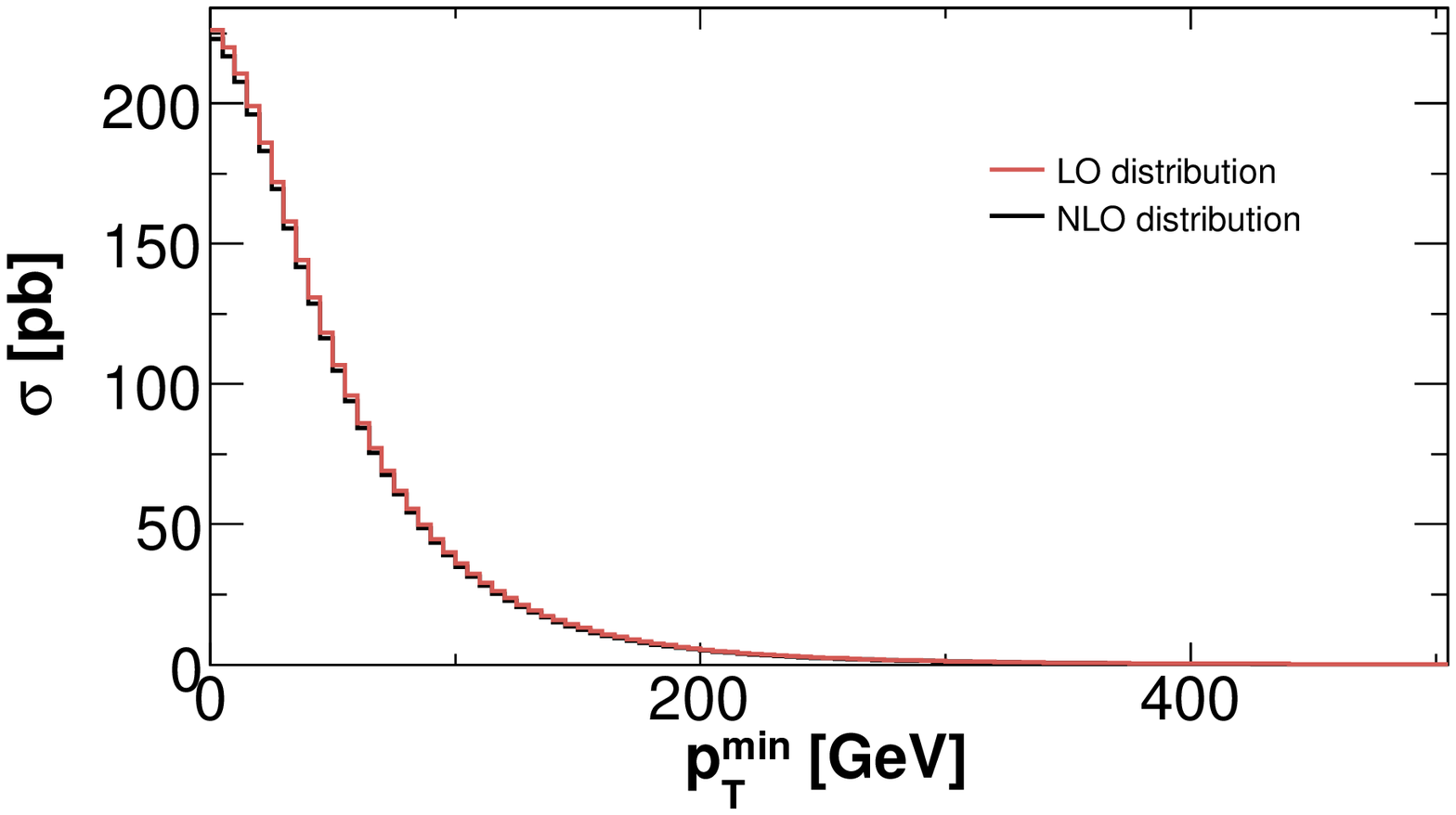  , width=8cm} 
\epsfig{file=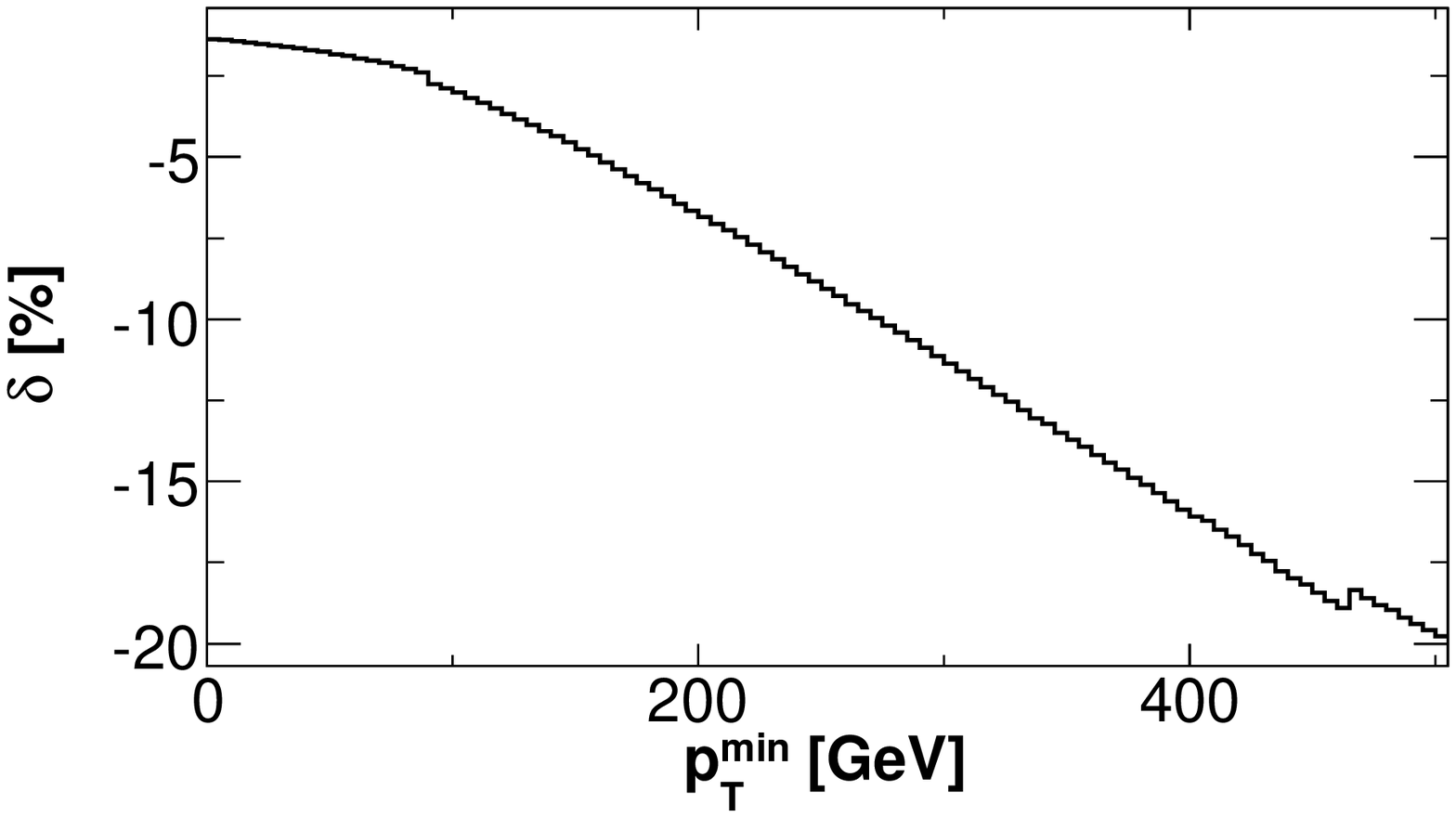, width=8cm} 

{\hspace{1cm} {\bf (c)} \hspace{8cm} {\bf (d)}}\\

\epsfig{file=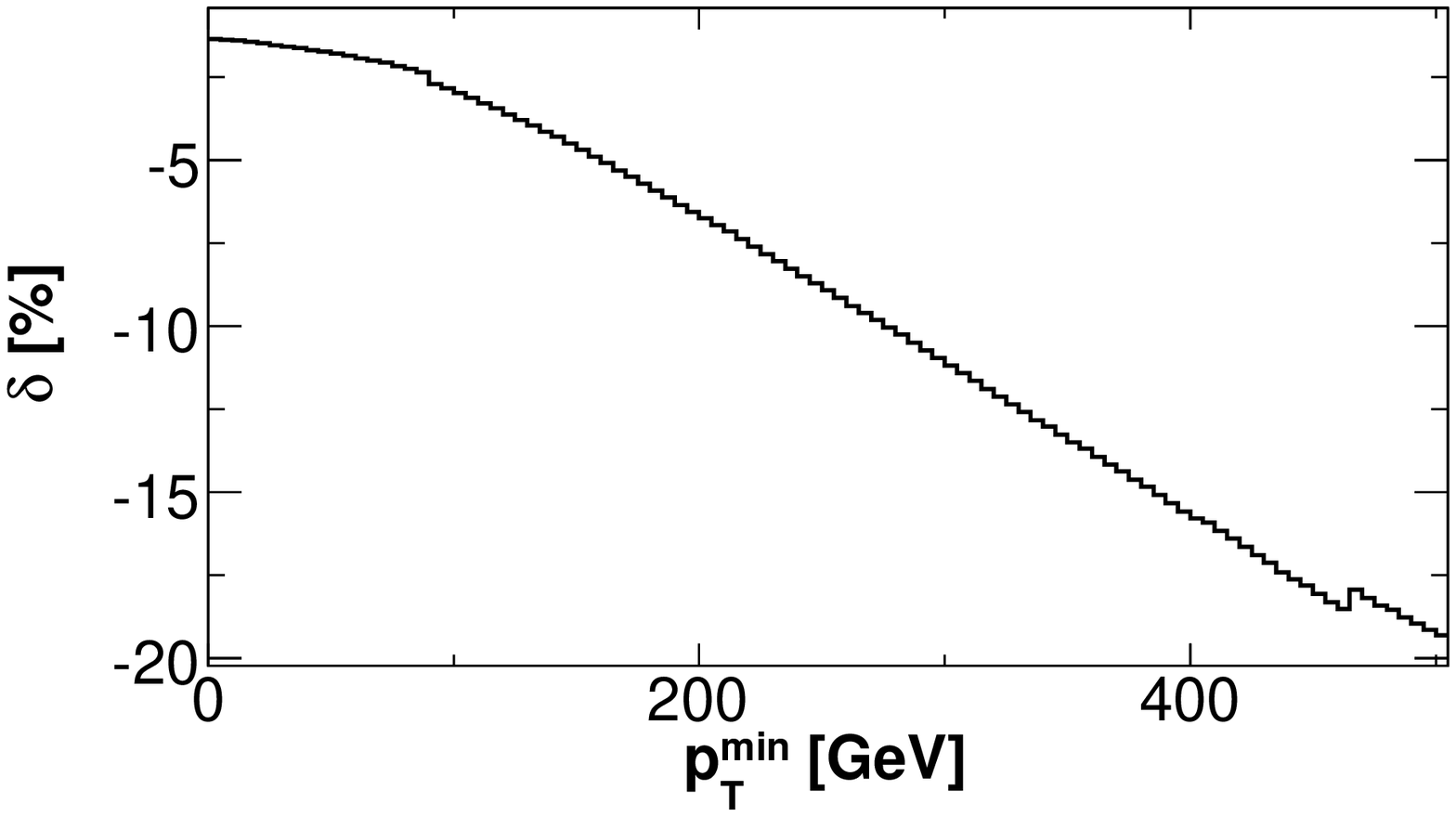, width=8cm} 
\epsfig{file=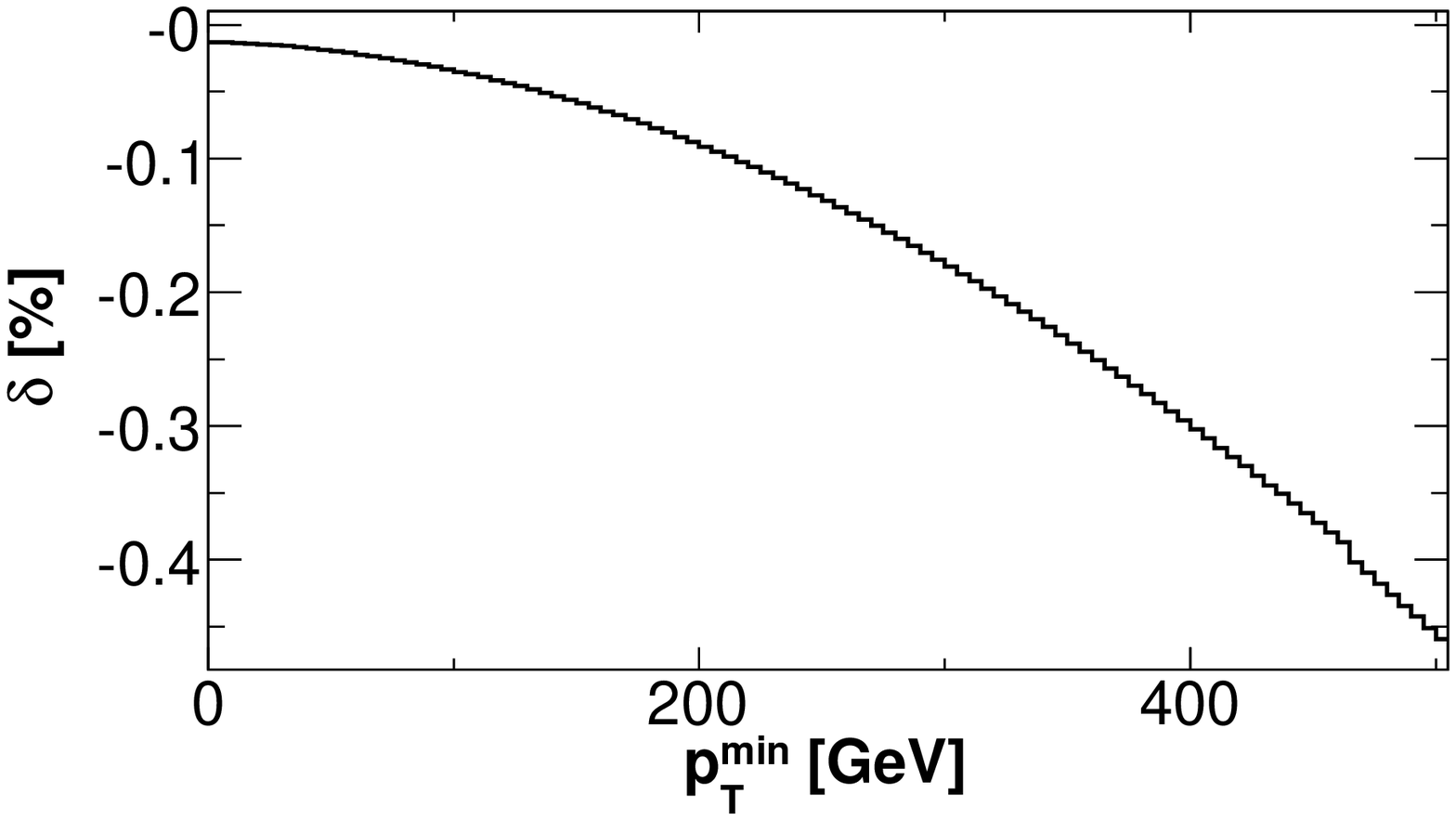, width=8cm} 
\caption{ \footnotesize{
{\bf (a)} We plot the LO (that is tree level)
contribution and the NLO that is tree level plus 
$\mathcal{O}(\alpha^3)$ plus SUSY QCD  
corrections to the integrated transverse momentum 
distribution $\sigma(p_T^{\rm min})$.
%
{\bf (b)} We plot the percentage contribution of 
the $\mathcal{O}(\alpha^3)$ plus SUSY QCD   corrections to the integrated transverse momentum 
distribution; that is $\delta = \frac{NLO -LO}{NLO}\times 100$.\\ 
%
%
{\bf (c)} We plot the percentage contribution of 
the $\mathcal{O}(\alpha^3)$   corrections to the integrated transverse momentum 
distribution; that is $\delta = \frac{\mathcal{O}(\alpha^3)}{NLO}\times 100$.\\ 
%
{\bf (d)} We plot the percentage contribution of 
the SUSY QCD   corrections to the integrated transverse momentum 
distribution; that is $\delta = \frac{SUSY~QCD}{NLO}\times 100$.\\  
${}$\\ 
No cuts are imposed.Computation in the SU1 point}}
\label{Fig:pT_int_SU1}
\end{figure}

\begin{figure}[htb] 
\centering 
{}
{\hspace{1cm} {\bf (a)} \hspace{8cm} {\bf (b)}}\\

\epsfig{file=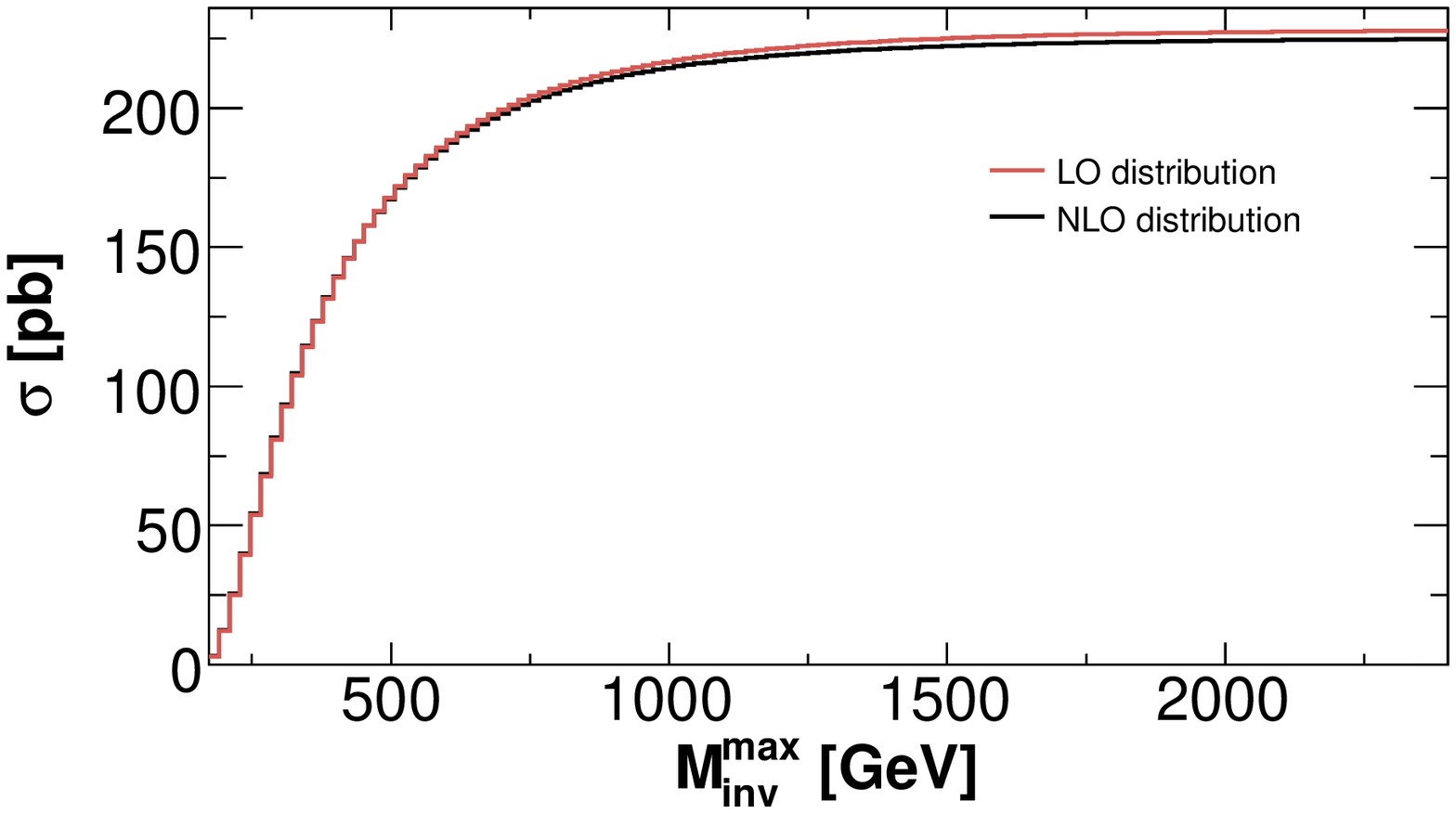 , width=8cm}
\epsfig{file=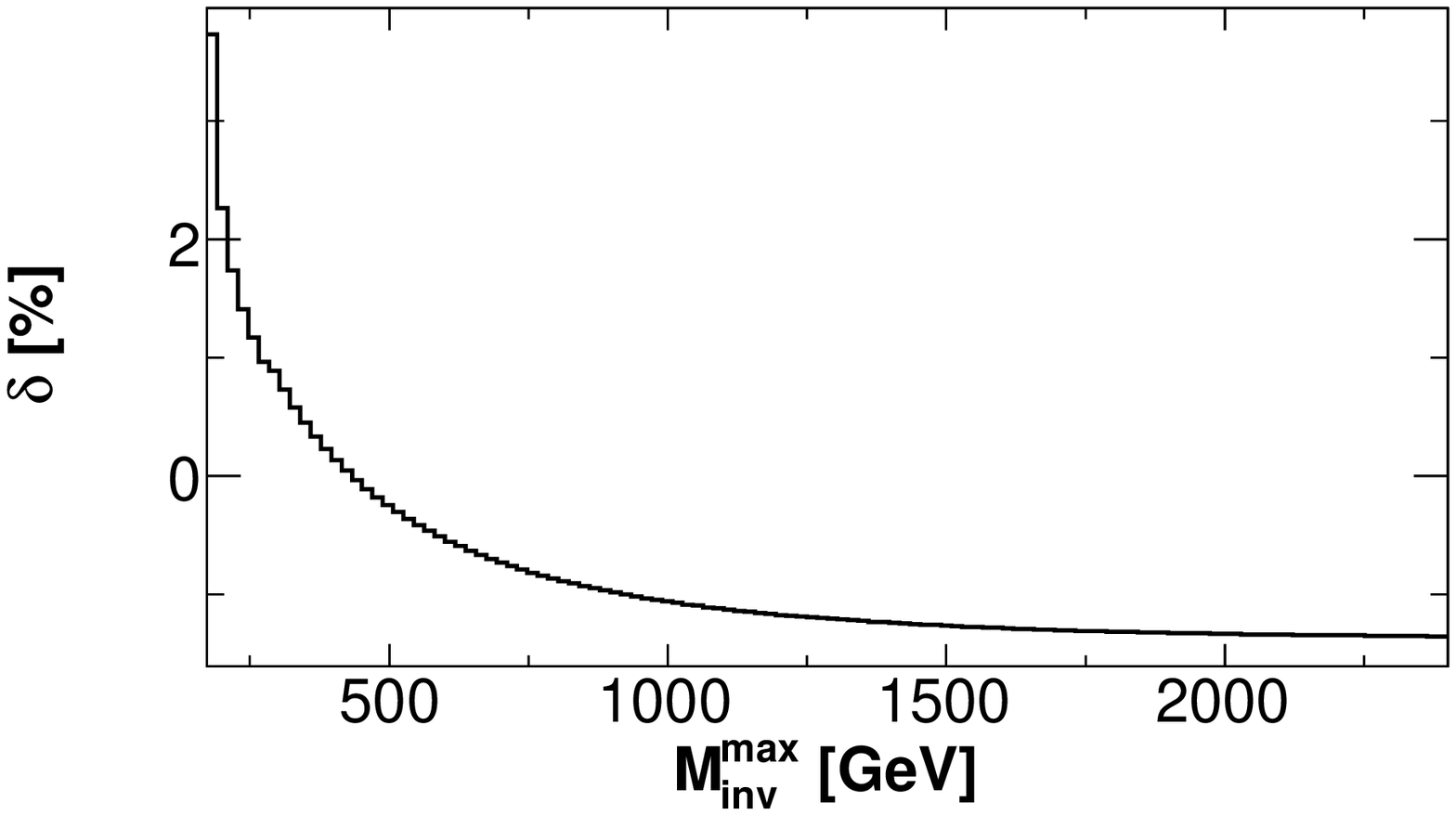, width=8cm} 

{\hspace{1cm} {\bf (c)} \hspace{8cm} {\bf (d)}}\\ 

\epsfig{file=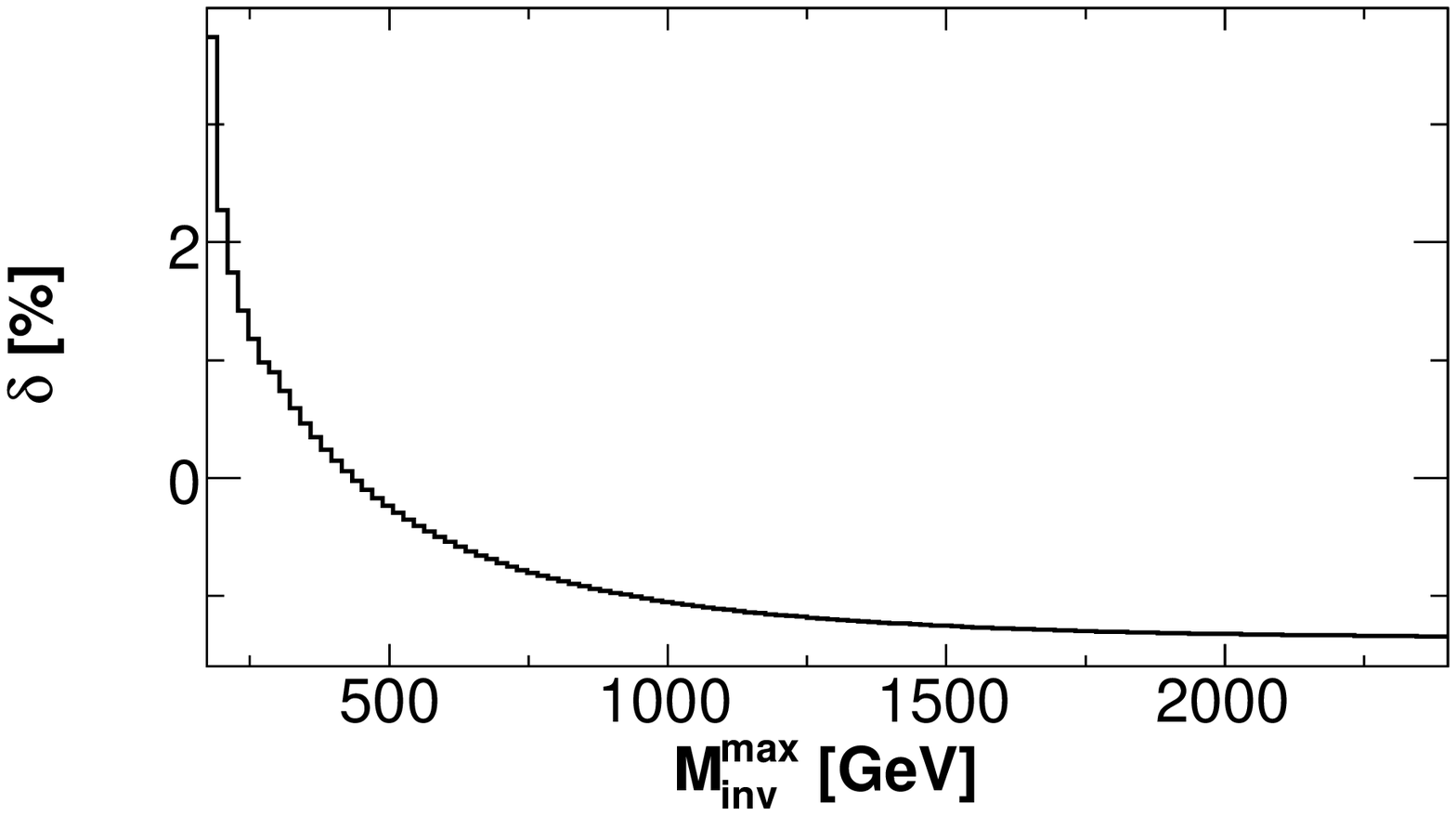, width=8cm} 
\epsfig{file=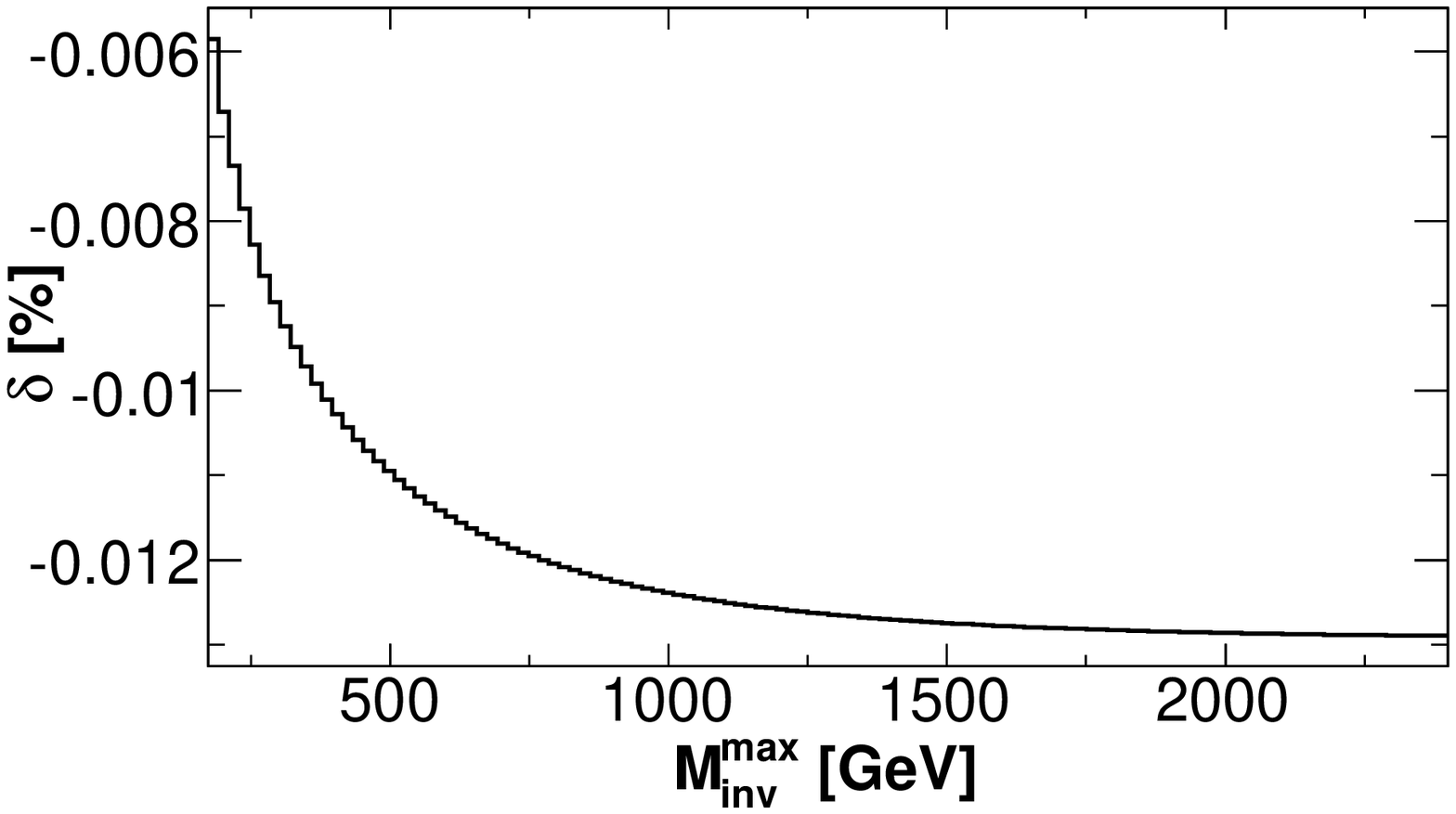, width=8cm} 

\caption{\footnotesize{
{\bf (a)} We plot the LO (that is tree level) 
contribution and the NLO (that is tree level plus $\mathcal{O}(\alpha^3)$ plus SUSY QCD) 
corrections to the cumulative invariant mass distribution $\sigma(M^{\rm max}_{\mbox{\tiny inv}})$. 
%
{\bf (b)} We plot the percentage contribution of  
the $\mathcal{O}(\alpha^3)$  plus SUSY QCD corrections to the cumulative invariant mass   
distribution; that is $\delta = \frac{NLO -LO}{NLO}\times 100$.\\  
%
{\bf (c)} We plot the percentage contribution of  
the $\mathcal{O}(\alpha^3)$ corrections to the cumulative invariant mass   
distribution; that is $\delta = \frac{\mathcal{O}(\alpha^3)}{NLO}\times 100$.\\  
%
{\bf (d)} We plot the percentage contribution of  
the SUSY QCD corrections to the cumulative invariant mass   
distribution; that is $\delta = \frac{SUSY~QCD}{NLO}\times 100$.\\  
${}$\\
No cuts are imposed. Computation in the SU1 point
}}  
\label{Fig:M_int_SU1} 
\end{figure}

\begin{figure}[htb] 
\centering 
{}
{\hspace{1cm} {\bf (a)} \hspace{8cm} {\bf (b)}}\\

\epsfig{file=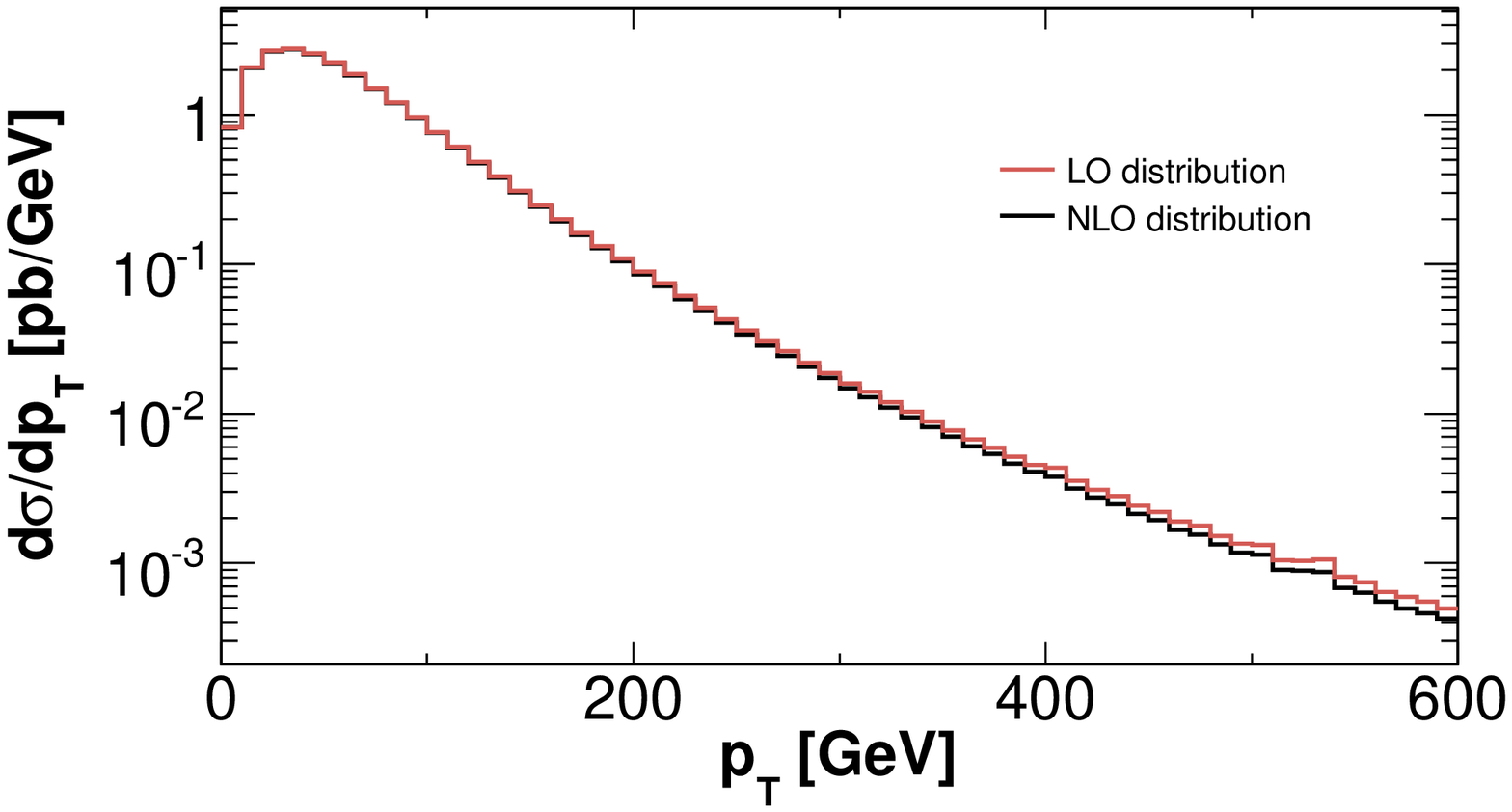 , width=8cm}
\epsfig{file=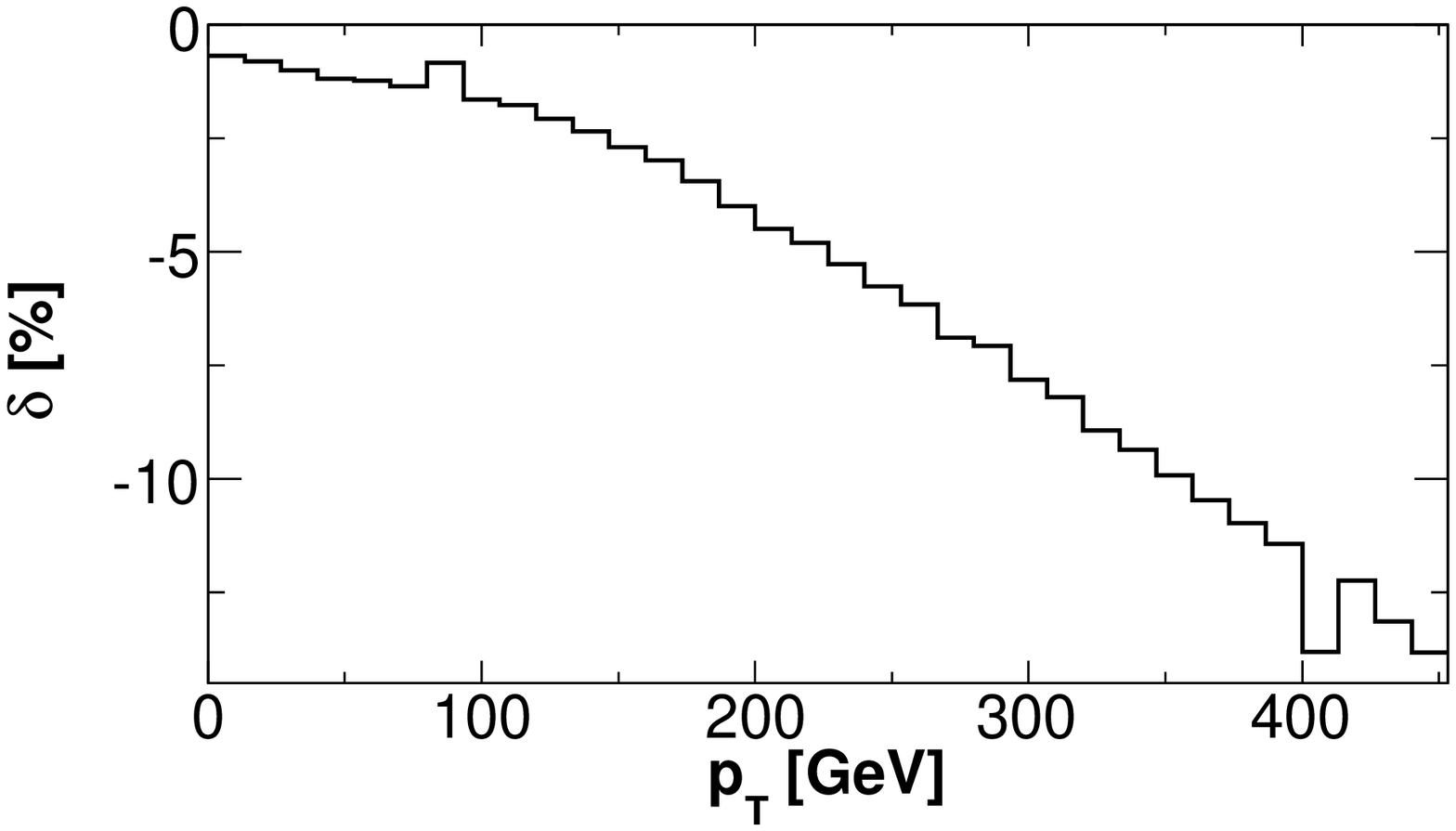, width=8cm}

{\hspace{1cm} {\bf (c)} \hspace{8cm} {\bf (d)}}\\

\epsfig{file=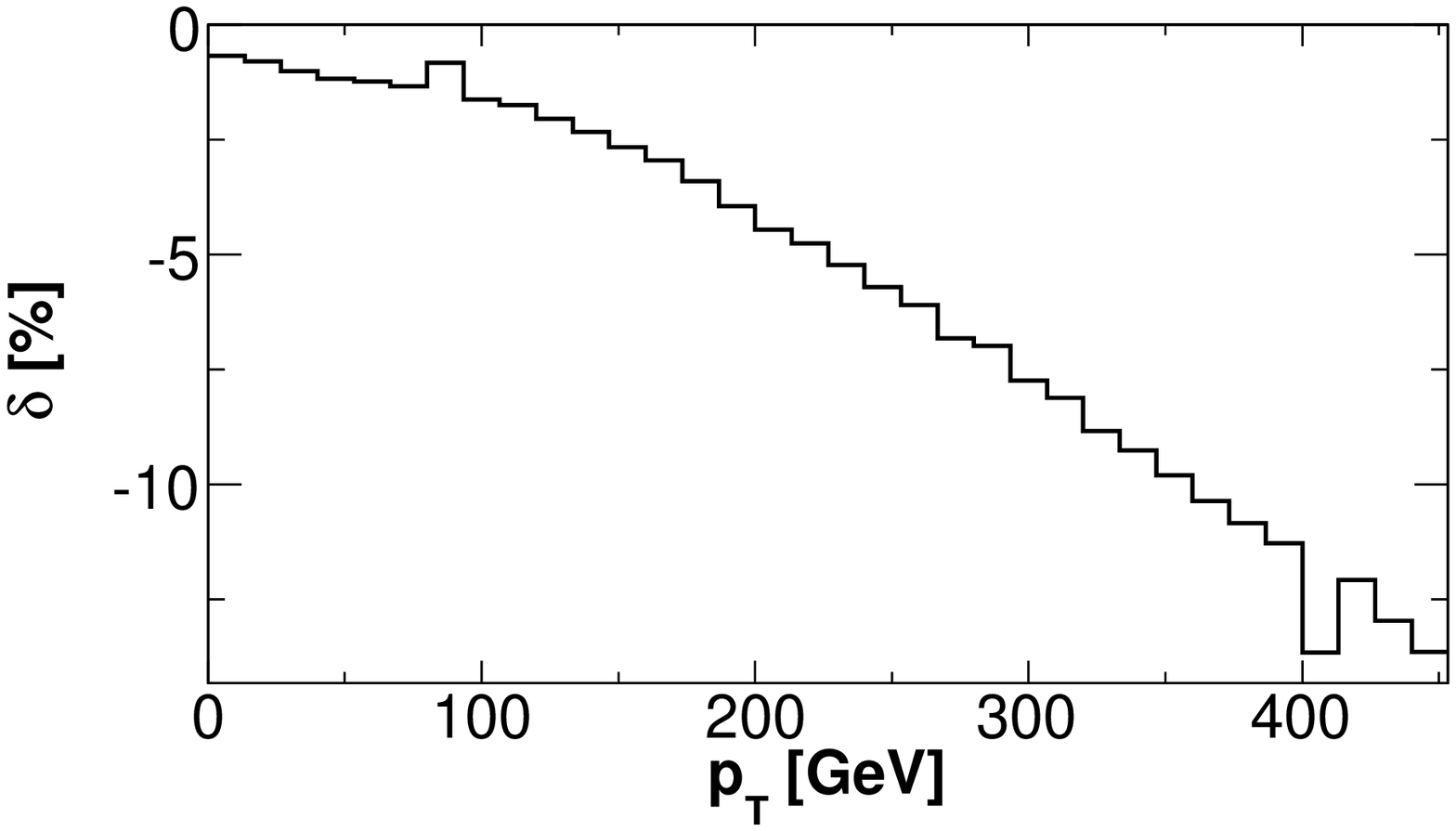, width=8cm} 
\epsfig{file=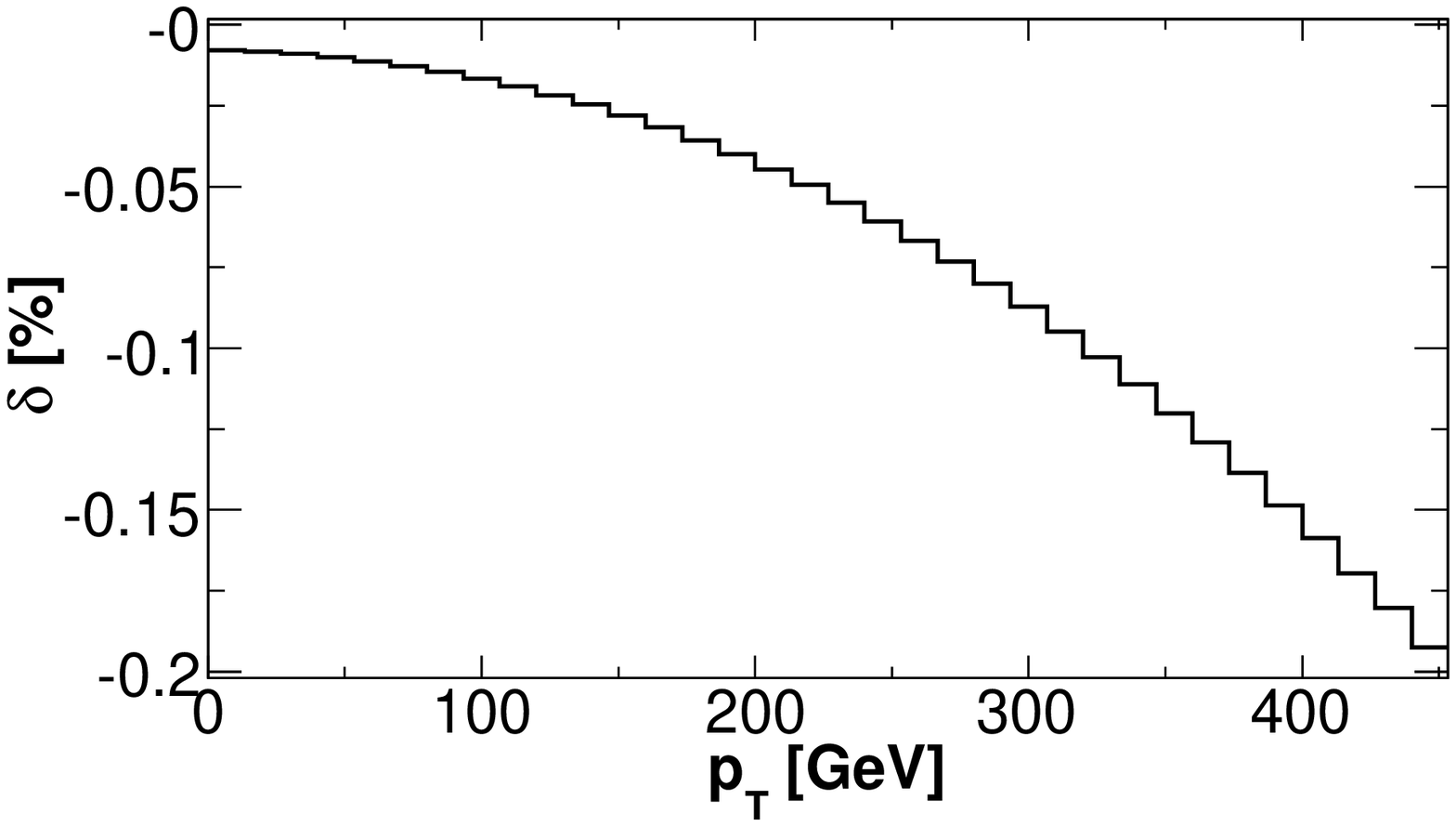, width=8cm}
\caption{\footnotesize{
{\bf(a) } We plot the LO (that is tree level) 
contribution and the NLO (that is tree level plus  
$\mathcal{O}(\alpha^3)$ plus SUSY QCD)   
corrections to the transverse momentum  
distribution.\\ 
%
{\bf (b)} We plot the percentage contribution of  
the $\mathcal{O}(\alpha^3)$ plus SUSY QCD   corrections to the transverse momentum  
distribution; that is $\delta = \frac{NLO -LO}{NLO}\times 100$.\\  
%
{\bf (c)} We plot the percentage contribution of  
the $\mathcal{O}(\alpha^3)$   corrections to the transverse momentum  
distribution; that is $\delta = \frac{\mathcal{O}(\alpha^3)}{NLO}\times 100$.\\  
%
{\bf (d)} We plot the percentage contribution of  
the SUSY QCD   corrections to the transverse momentum  
distribution; that is $\delta = \frac{SUSY~QCD}{NLO}\times 100$.\\  
${}$\\
No cuts are imposed. Computation in the SU6 point}
}  
\label{Fig:pT_SU6} 
\end{figure}

\begin{figure}[htb] 
\centering 
{}
{\hspace{1cm} {\bf (a)} \hspace{8cm} {\bf (b)}}\\

\epsfig{file=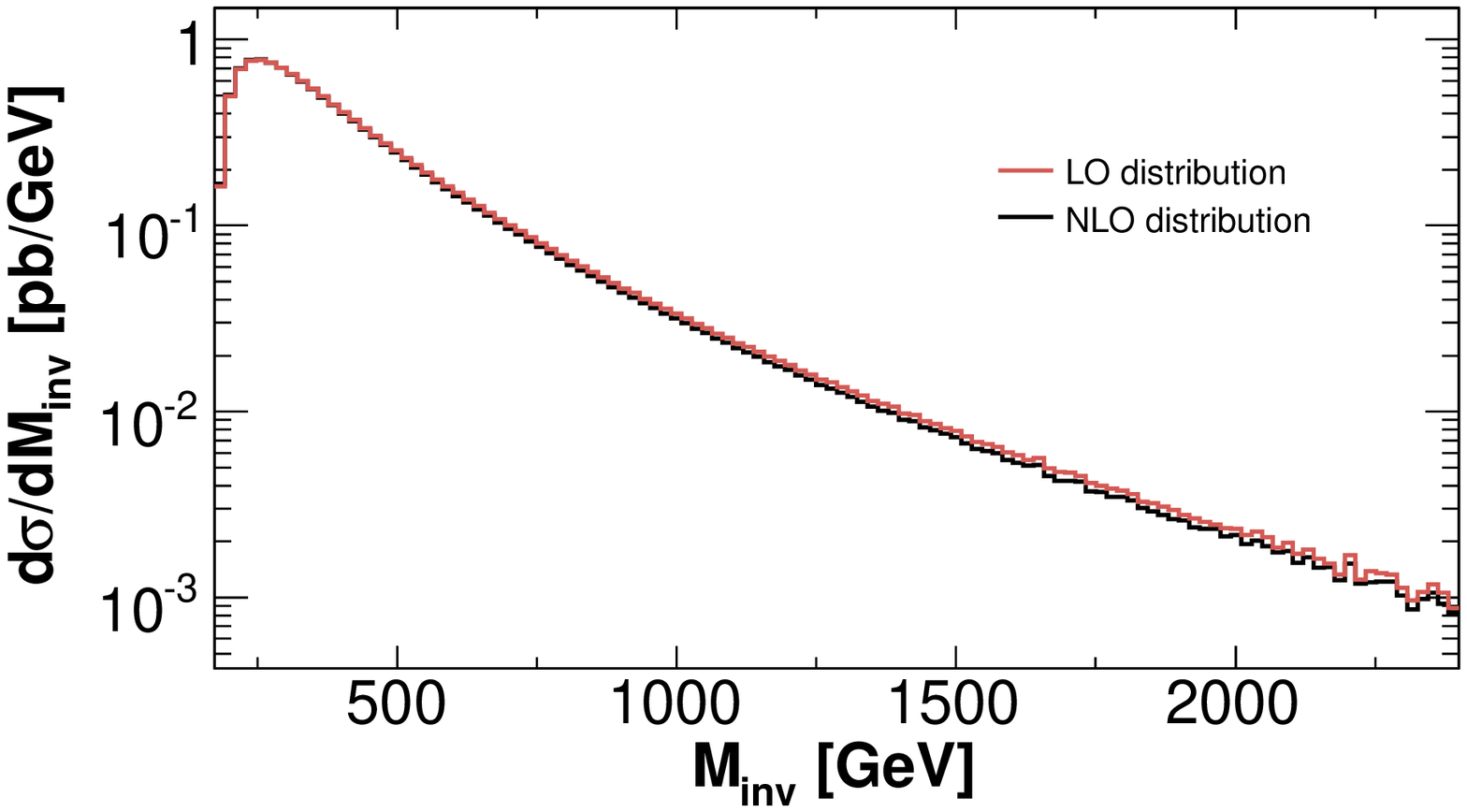 , width=8cm} 
\epsfig{file=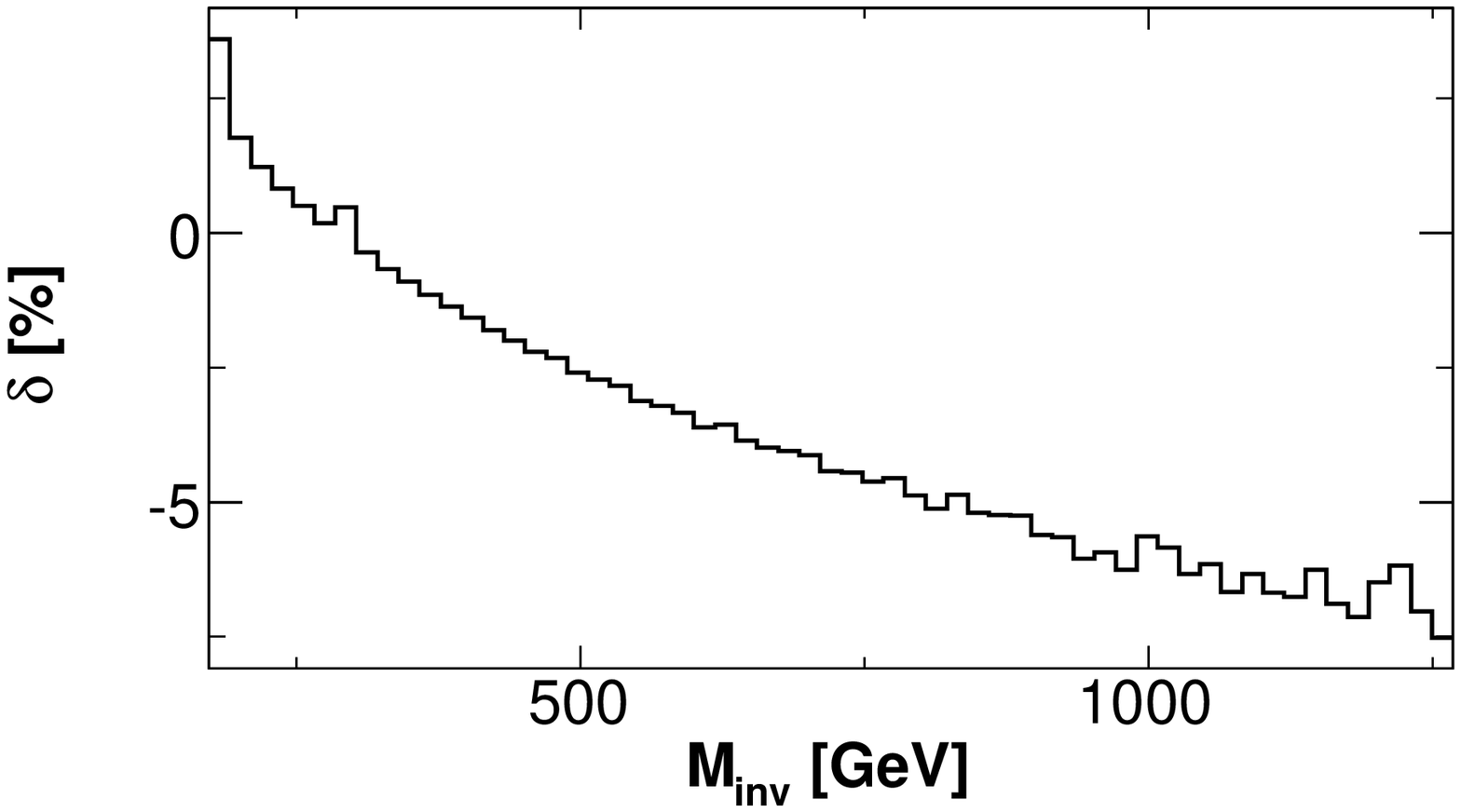, width=8cm} 

{\hspace{1cm} {\bf (c)} \hspace{8cm} {\bf (d)}}\\

\epsfig{file=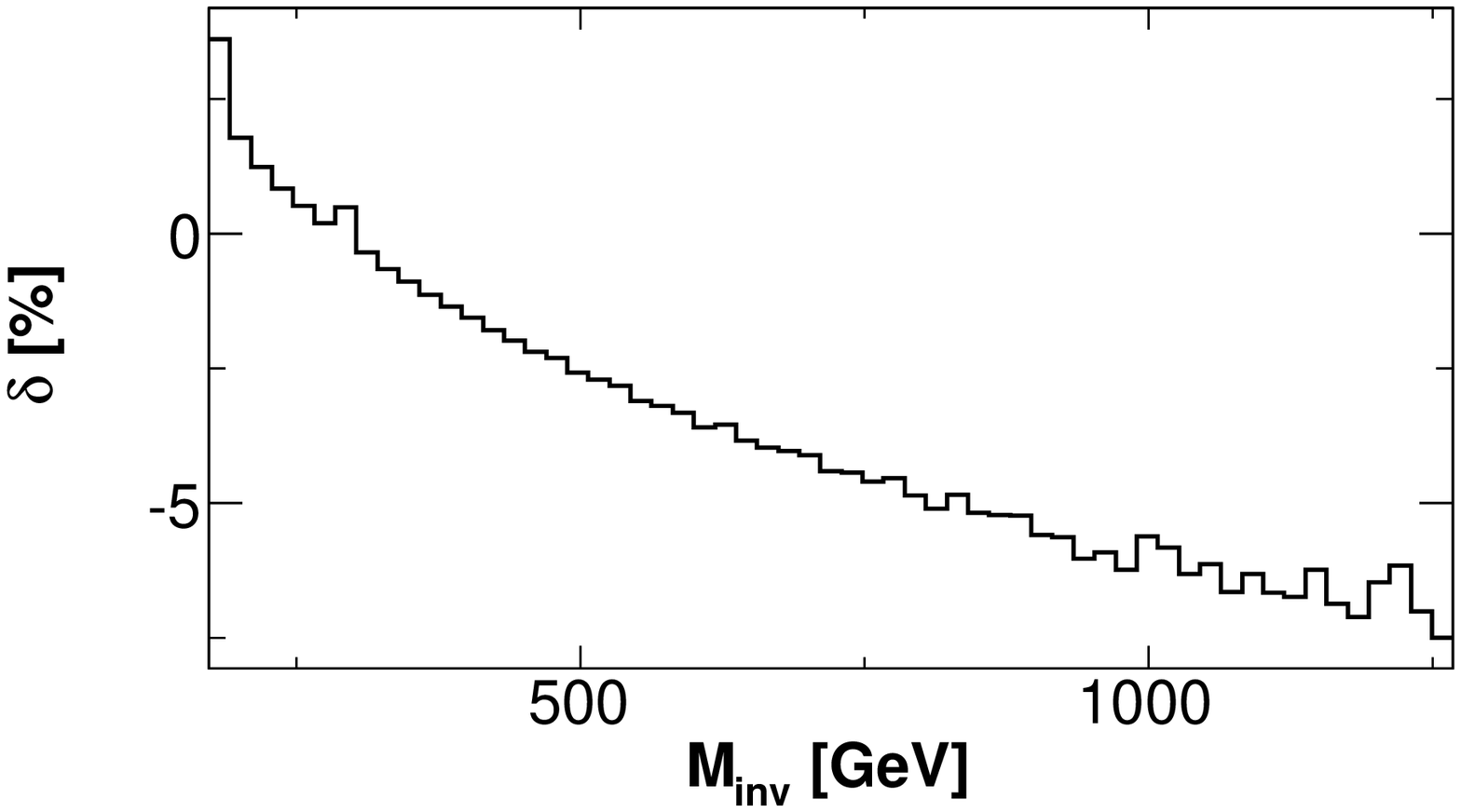, width=8cm} 
\epsfig{file=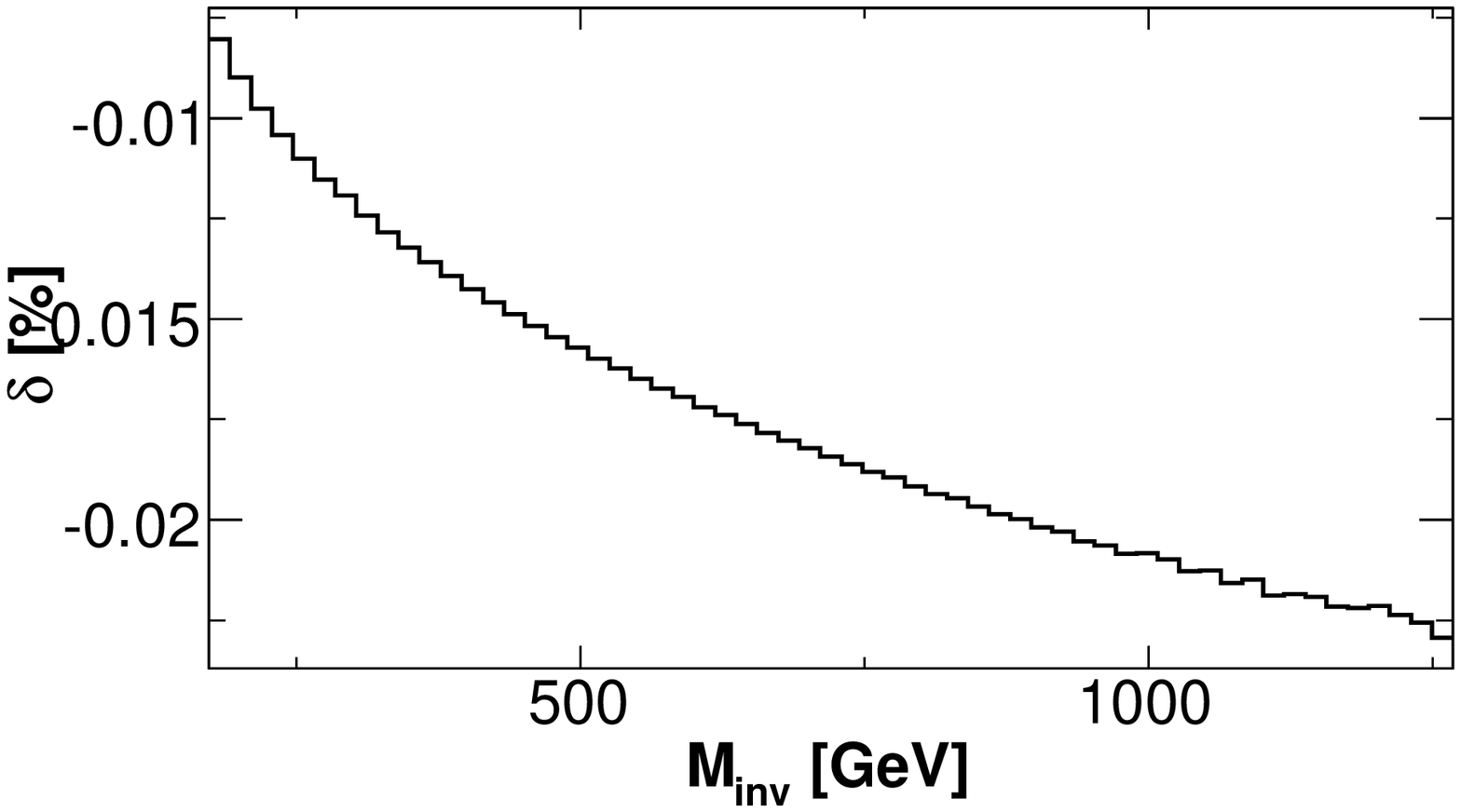, width=8cm} 
\caption{\footnotesize{
{\bf (a)} We plot the LO (that is tree level) 
contribution and the NLO (that is tree level plus $\mathcal{O}(\alpha^3)$ plus SUSY QCD) 
corrections to the invariant mass distribution.\\ 
%
{\bf (b)} We plot the percentage contribution of  
the $\mathcal{O}(\alpha^3)$ plus SUSY QCD   corrections to the invariant mass   
distribution; that is $\delta = \frac{NLO -LO}{NLO}\times 100$.\\  
%
{\bf (c)} We plot the percentage contribution of  
the $\mathcal{O}(\alpha^3)$   corrections to the invariant mass   
distribution; that is $\delta = \frac{\mathcal{O}(\alpha^3)}{NLO}\times 100$.\\  
%
{\bf (d)} We plot the percentage contribution of  
the SUSY QCD   corrections to the invariant mass   
distribution; that is $\delta = \frac{SUSY~QCD}{NLO}\times 100$.\\  
${}$\\
No cuts are imposed. Computation in the SU6 point
}} 
\label{Fig:M_distr_SU6} 
\end{figure}

\begin{figure}
\centering
{}
{\hspace{1cm} {\bf (a)} \hspace{8cm} {\bf (b)}}\\

\epsfig{file=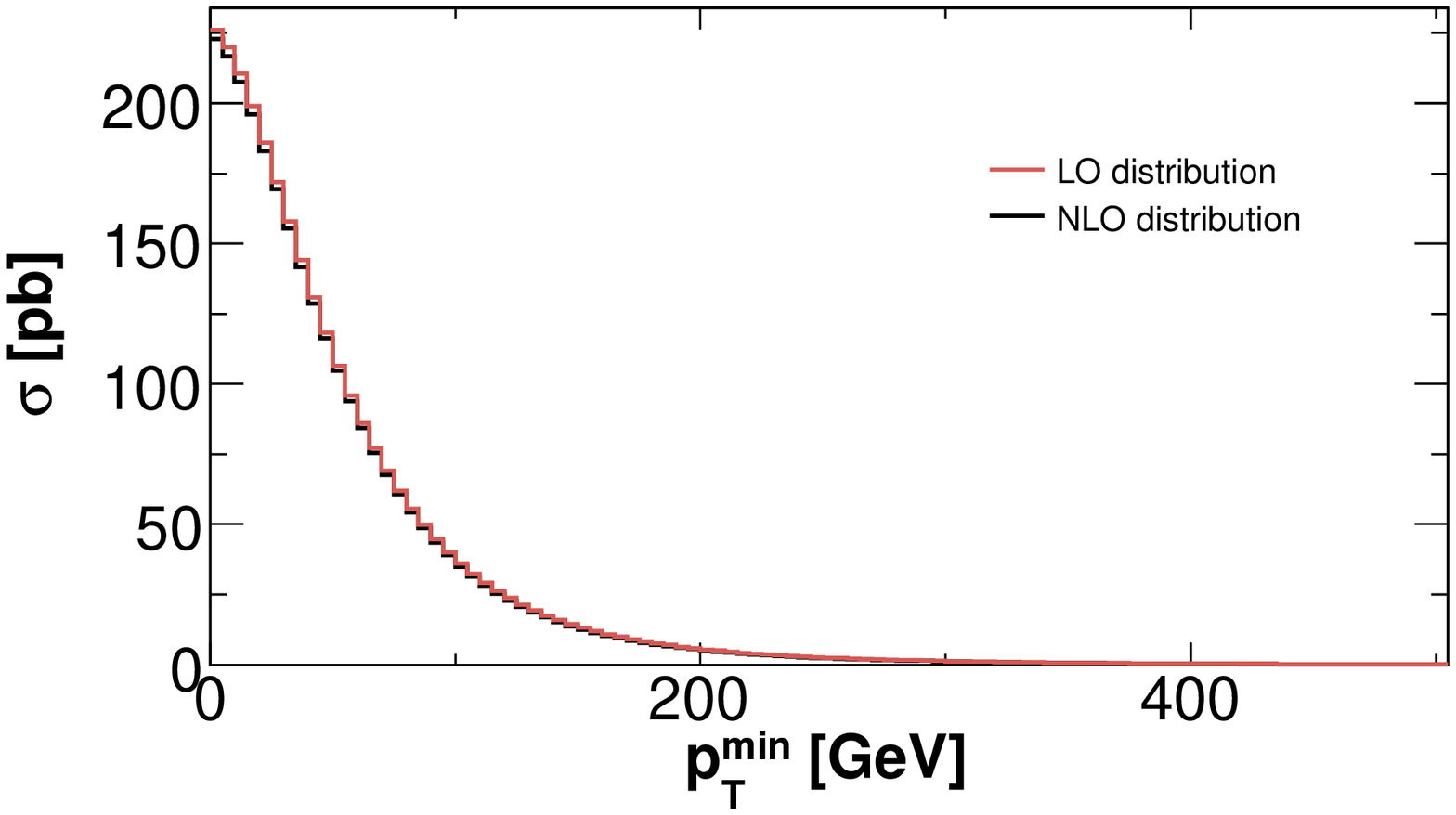  , width=8cm} 
\epsfig{file=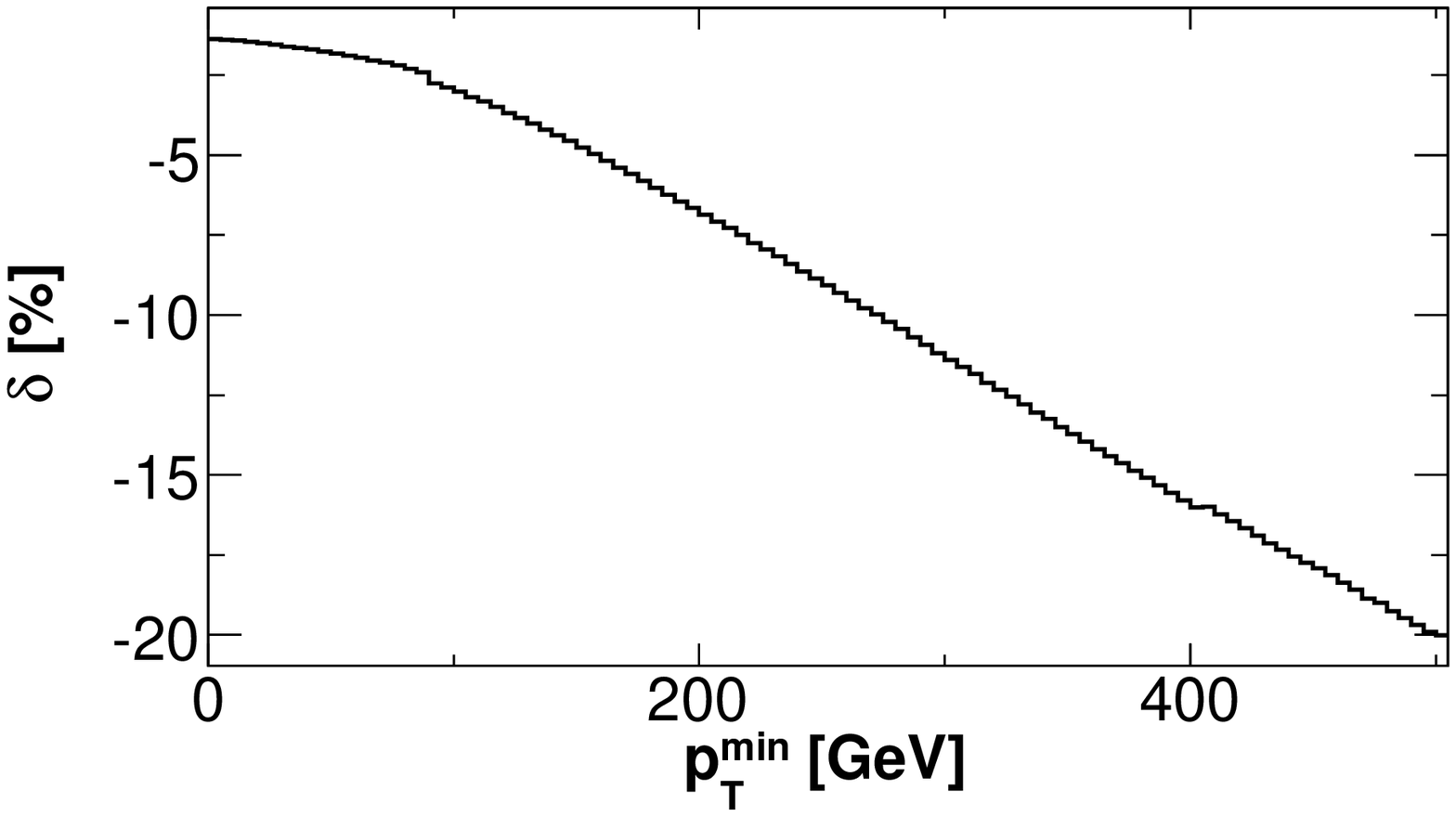, width=8cm} 

{\hspace{1cm} {\bf (c)} \hspace{8cm} {\bf (d)}}\\

\epsfig{file=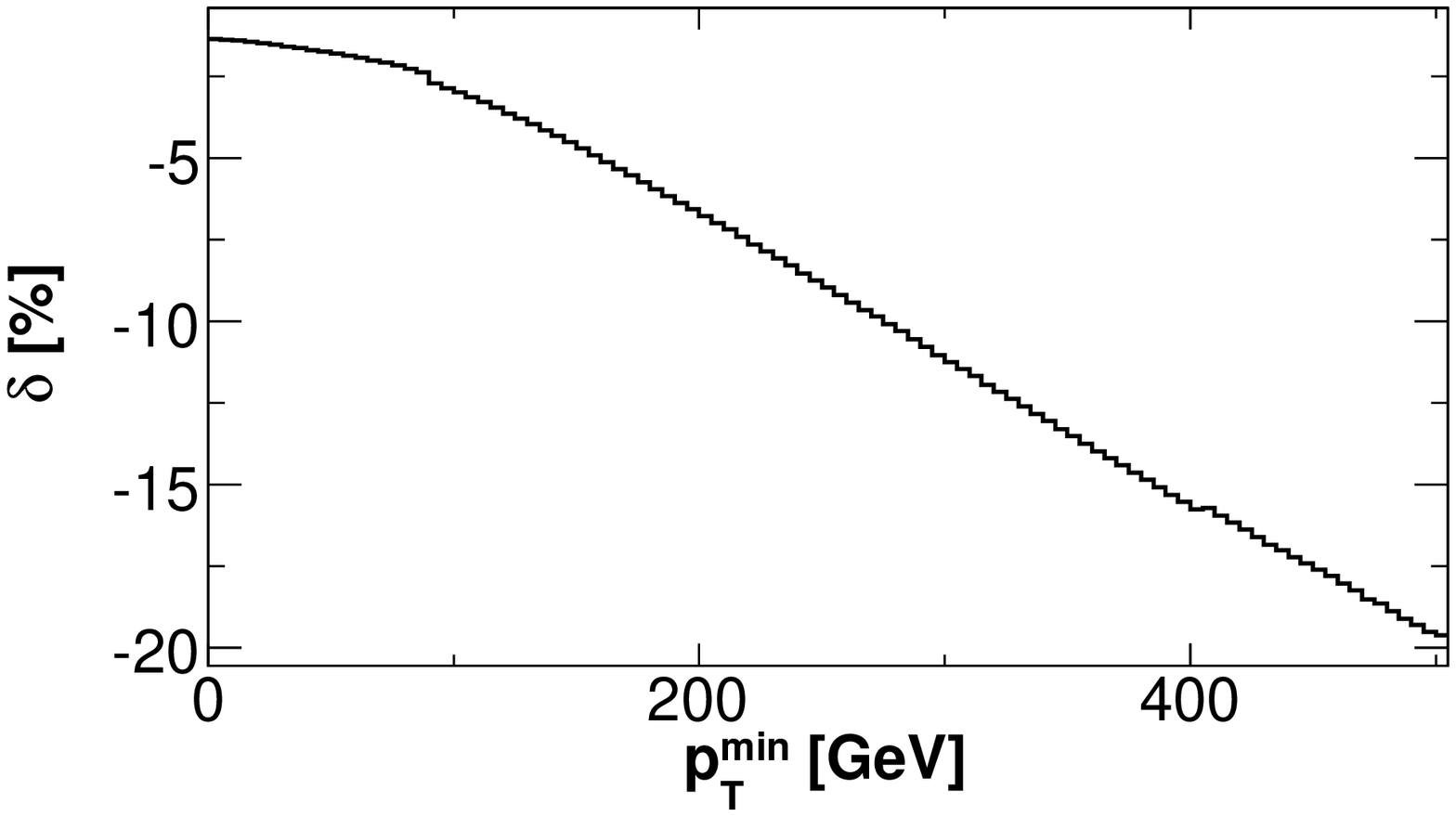, width=8cm} 
\epsfig{file=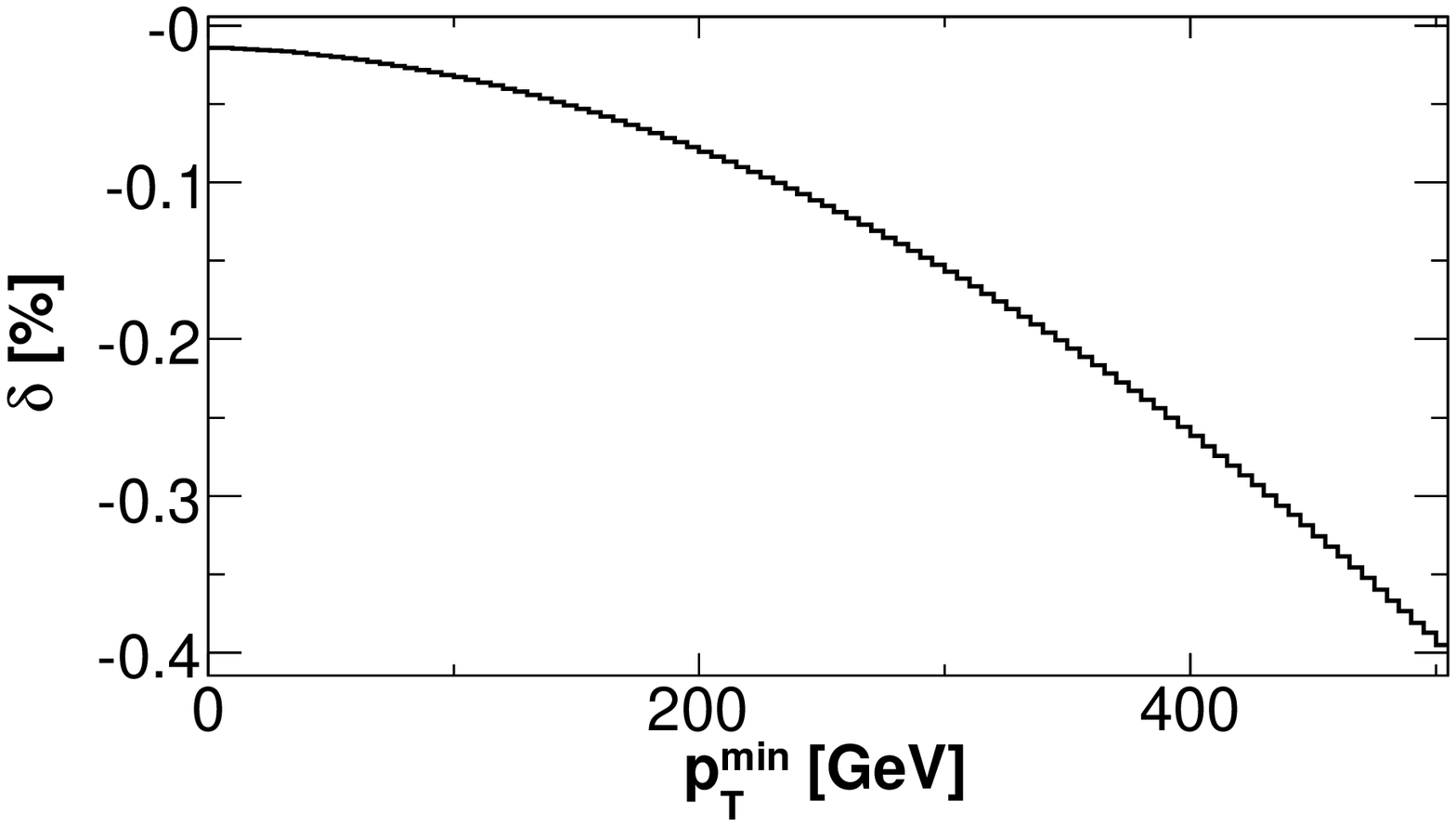, width=8cm} 
\caption{ \footnotesize{
{\bf (a)} We plot the LO (that is tree level)
contribution and the NLO that is tree level plus 
$\mathcal{O}(\alpha^3)$ plus SUSY QCD  
corrections to the integrated transverse momentum 
distribution $\sigma(p_T^{\rm min})$.
%
{\bf (b)} We plot the percentage contribution of 
the $\mathcal{O}(\alpha^3)$ plus SUSY QCD   corrections to the integrated transverse momentum 
distribution; that is $\delta = \frac{NLO -LO}{NLO}\times 100$.\\ 
%
%
{\bf (c)} We plot the percentage contribution of 
the $\mathcal{O}(\alpha^3)$   corrections to the integrated transverse momentum 
distribution; that is $\delta = \frac{\mathcal{O}(\alpha^3)}{NLO}\times 100$.\\ 
%
{\bf (d)} We plot the percentage contribution of 
the SUSY QCD   corrections to the integrated transverse momentum 
distribution; that is $\delta = \frac{SUSY~QCD}{NLO}\times 100$.\\  
${}$\\ 
No cuts are imposed.Computation in the SU6 point}}
\label{Fig:pT_int_SU6}
\end{figure}

\begin{figure}[htb] 
\centering 
{}
{\hspace{1cm} {\bf (a)} \hspace{8cm} {\bf (b)}}\\

\epsfig{file=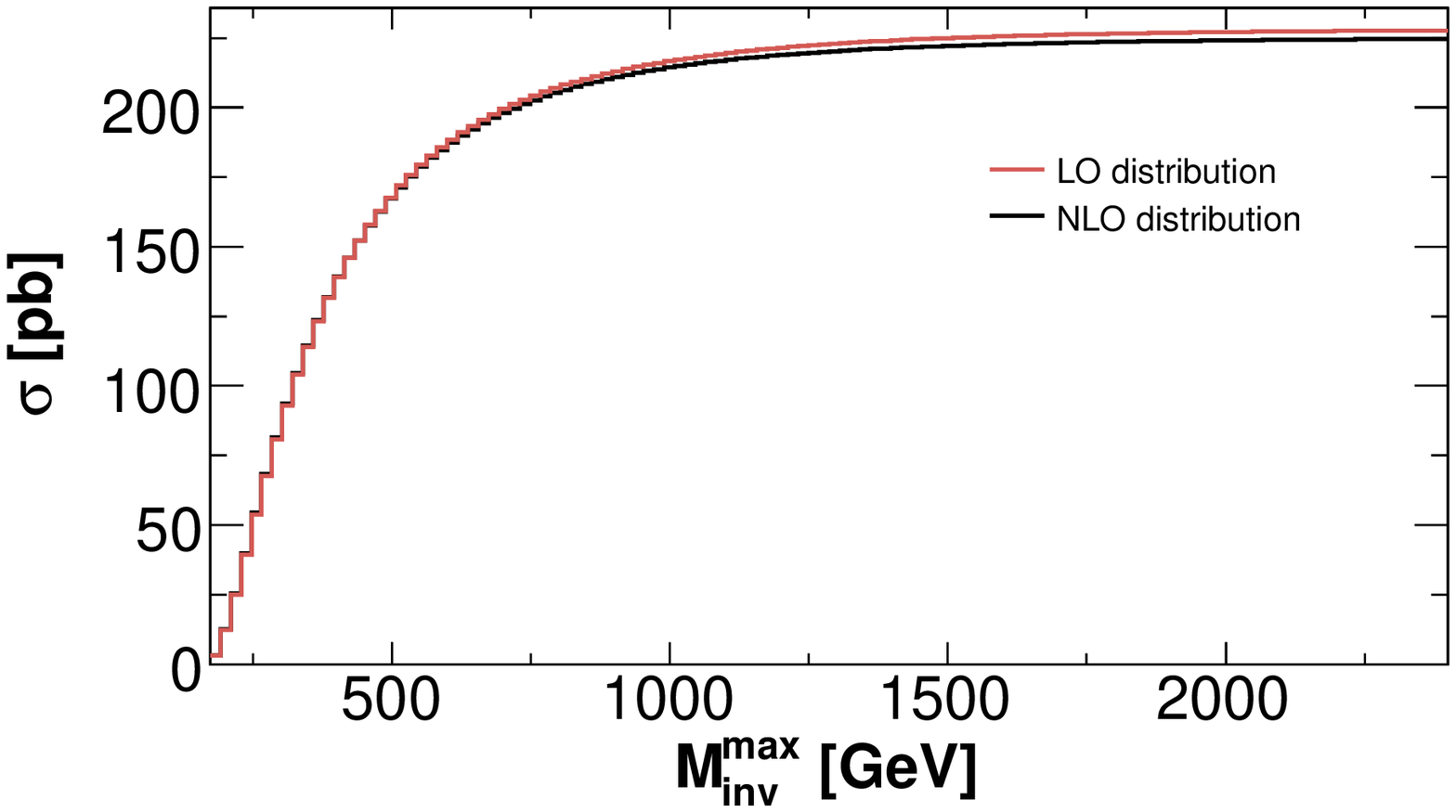 , width=8cm}
\epsfig{file=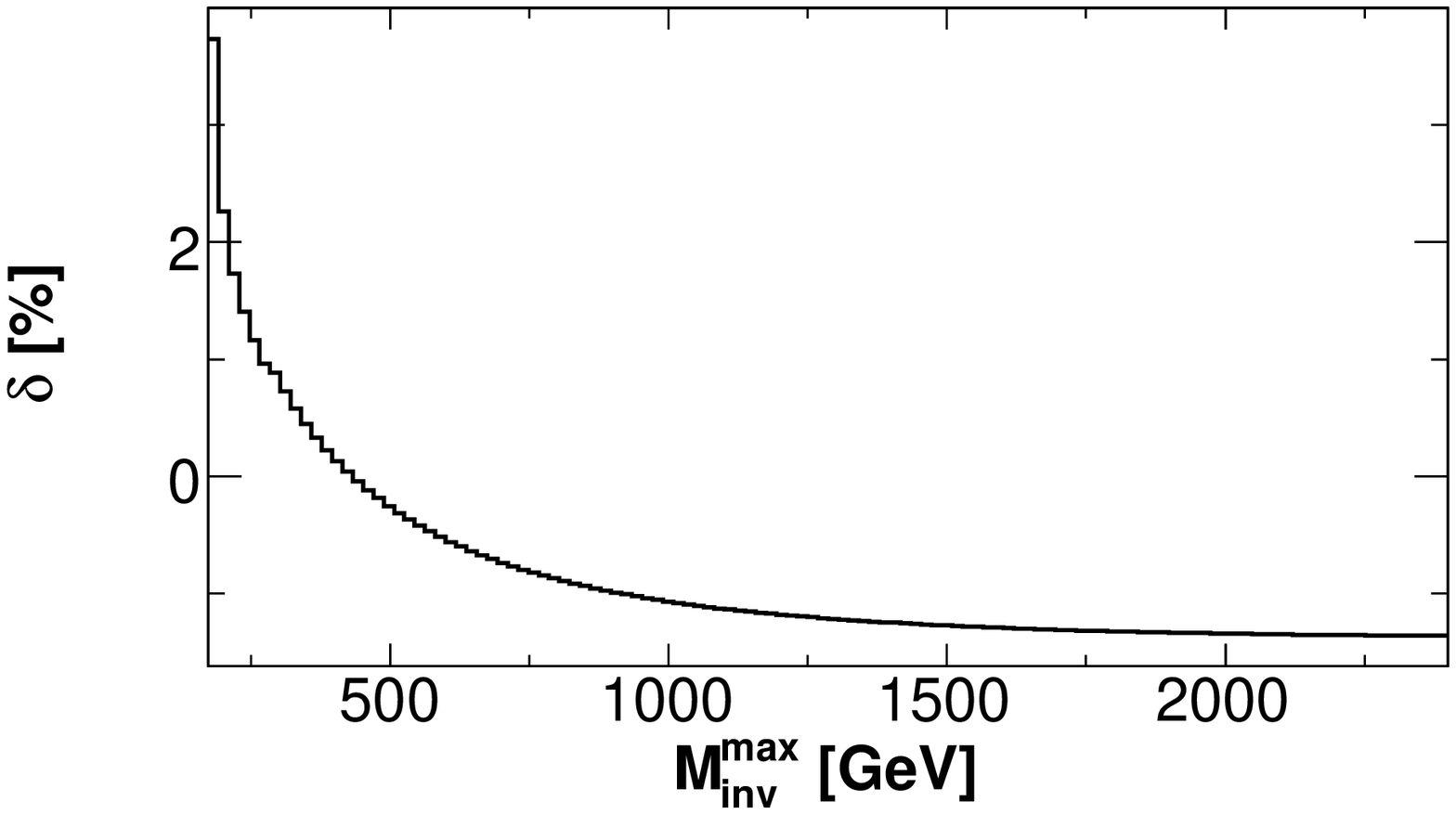, width=8cm} 

{\hspace{1cm} {\bf (c)} \hspace{8cm} {\bf (d)}}\\ 

\epsfig{file=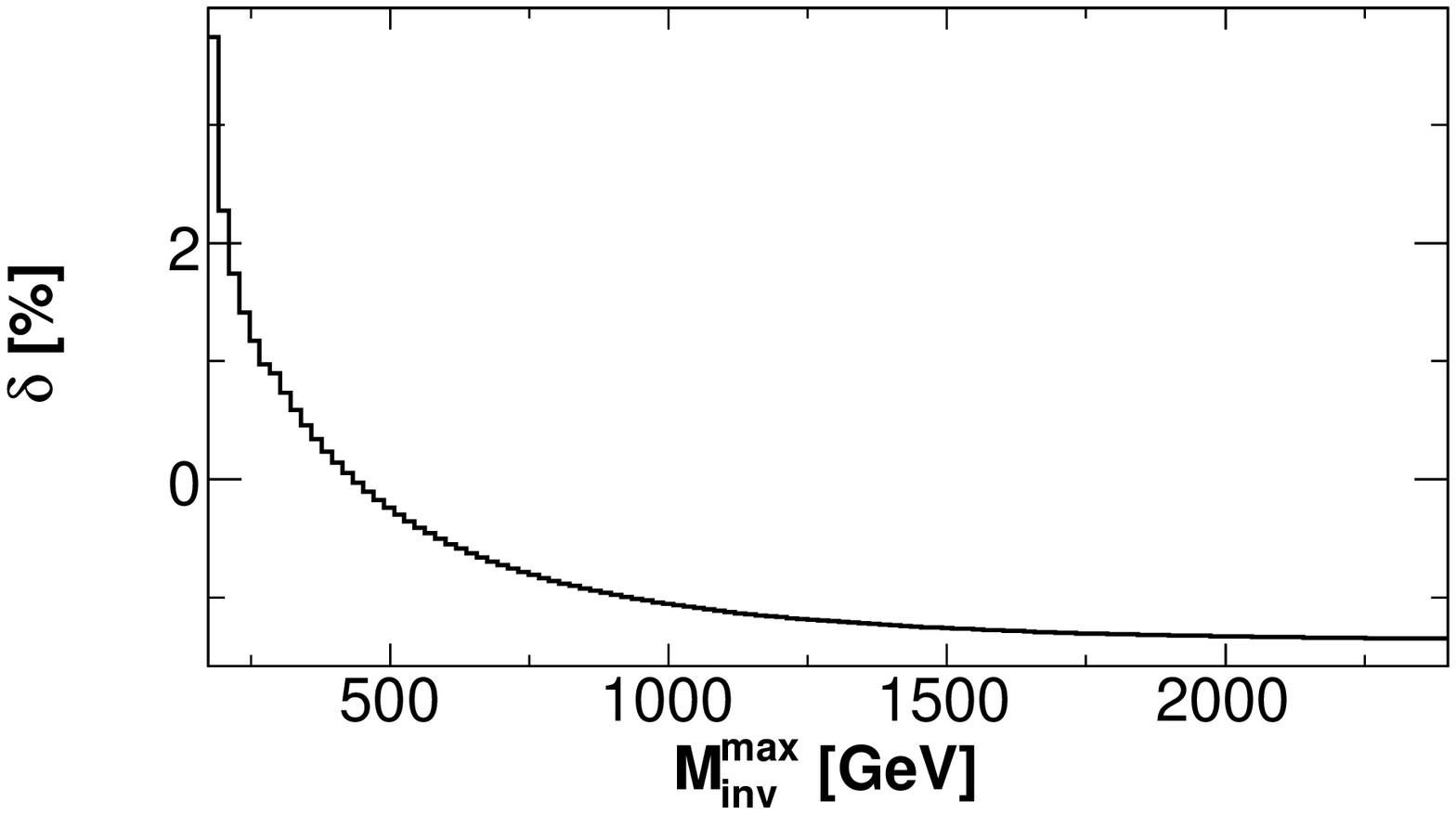, width=8cm} 
\epsfig{file=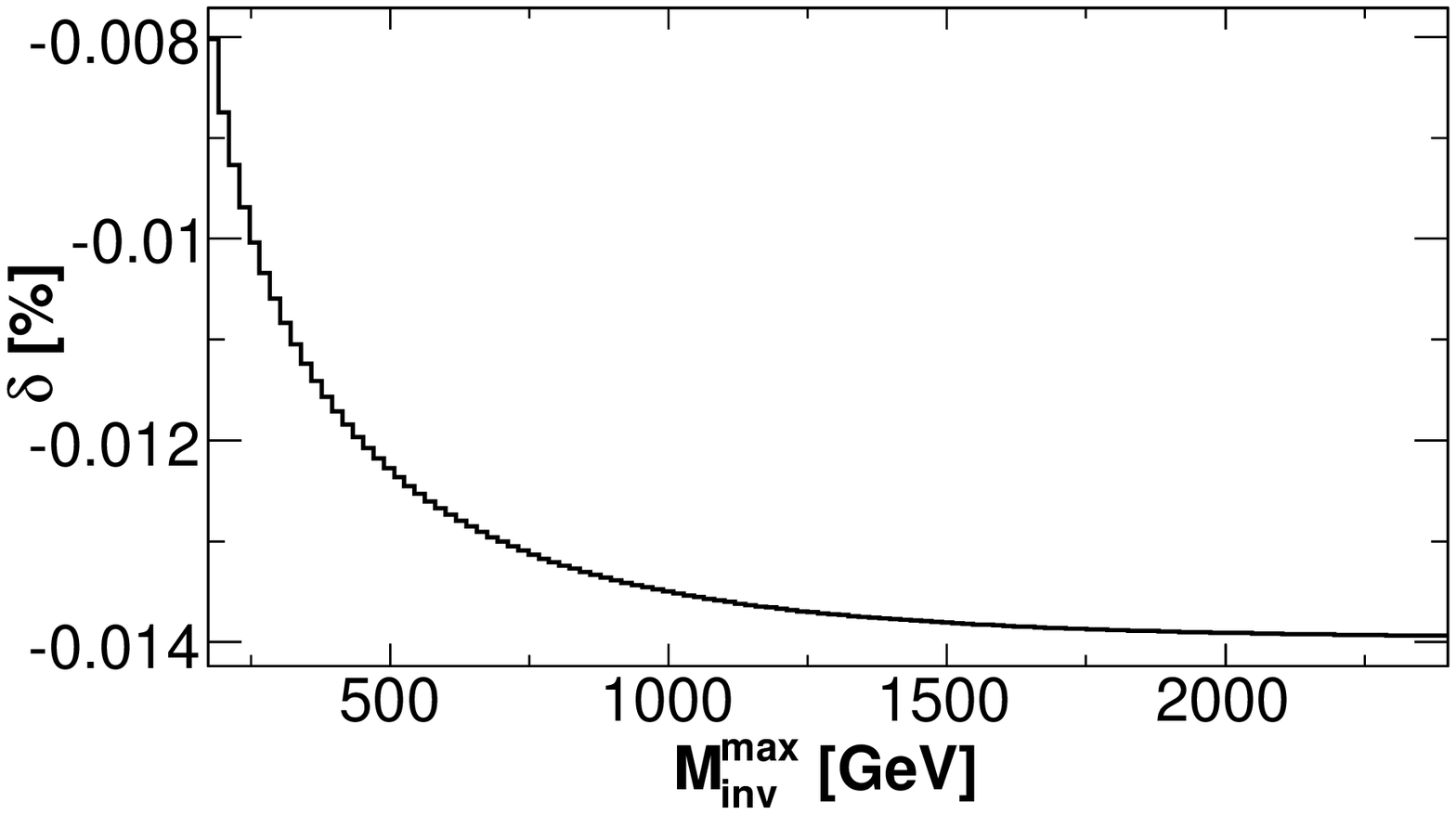, width=8cm} 

\caption{\footnotesize{
{\bf (a)} We plot the LO (that is tree level) 
contribution and the NLO (that is tree level plus $\mathcal{O}(\alpha^3)$ plus SUSY QCD) 
corrections to the cumulative invariant mass distribution $\sigma(M^{\rm max}_{\mbox{\tiny inv}})$. 
%
{\bf (b)} We plot the percentage contribution of  
the $\mathcal{O}(\alpha^3)$  plus SUSY QCD corrections to the cumulative invariant mass   
distribution; that is $\delta = \frac{NLO -LO}{NLO}\times 100$.\\  
%
{\bf (c)} We plot the percentage contribution of  
the $\mathcal{O}(\alpha^3)$ corrections to the cumulative invariant mass   
distribution; that is $\delta = \frac{\mathcal{O}(\alpha^3)}{NLO}\times 100$.\\  
%
{\bf (d)} We plot the percentage contribution of  
the SUSY QCD corrections to the cumulative invariant mass   
distribution; that is $\delta = \frac{SUSY~QCD}{NLO}\times 100$.\\  
${}$\\
No cuts are imposed. Computation in the SU6 point
}}  
\label{Fig:M_int_SU6} 
\end{figure}

The following figures show the analogous  
results for the transverse momentum distribution 
$\frac{d \sigma}{dp_{T}}$, the invariant mass distribution $\frac{d 
  \sigma}{dM_{\mbox{\tiny inv}}}$,  the integrated $p_T$ ditribution $\sigma(p_T^{\rm min})$ 
and the cumulative invariant mass distribution 
$\sigma(M^{\rm max}_{ \mbox{\tiny inv} })$ for the SU1  
(Fig.~\ref{Fig:pT_SU1},~\ref{Fig:M_distr_SU1},~\ref{Fig:pT_int_SU1},~\ref{Fig:M_int_SU1}) and for 
the SU6  (Fig.~\ref{Fig:pT_SU6},~\ref{Fig:M_distr_SU6},~\ref{Fig:pT_int_SU6},~\ref{Fig:M_int_SU6}) benchmark points.  
In the MSSM cases the NLO is defined as the sum of the electroweak part 
$\mathcal{O}(\alpha^3)$ and of the SUSY QCD part; in each figure we show in the panel 
{\bf (a)}  the behaviour of the observable at LO and NLO; 
in {\bf (b)}  the global NLO  effect (where "global" means $\mathcal{O}(\alpha^3)$ plus SUSY QCD); 
in {\bf(c)} and {\bf{(d)}} the separate percentage contributions of 
the $\mathcal{O}(\alpha^3)$ and SUSY QCD parts respectively.\\ 
A first comment that can be drawn from the inspection of the plots is that the 
$t$-channel process is very weakly sensitive to the presence of the 
supersymmetric particle in the loops. The difference with the Standard Model case, below 
the percent level, is negligible for both the mSUGRA benchmark points, without 
appreciable differences between the two sets.\\ 
The plots for percentage effect of the SUSY QCD corrections only  
(the {\bf (d)} panel in
Fig. ~\ref{Fig:pT_SU1},~\ref{Fig:M_distr_SU1},~\ref{Fig:pT_int_SU1},~\ref{Fig:M_int_SU1} for SU1 and ~
\ref{Fig:pT_SU6},~\ref{Fig:M_distr_SU6},~\ref{Fig:pT_int_SU6},~\ref{Fig:M_int_SU6} for SU6)
show that the contribution of diagrams with virtual squark and gluinos  
is systematically small, below the $1 \%$ level, in agreement  with Ref.~\cite{Zhang:2006cx}. 
The same conclusion holds for the pure electroweak supersymmetric contribution,  
being the $\mathcal{O}(\alpha^3)$ effect due almost completely to the Standard Model part.  
 
\subsubsection{PDF uncertainties}
\begin{figure}[htb] 
\centering 
\epsfig{file=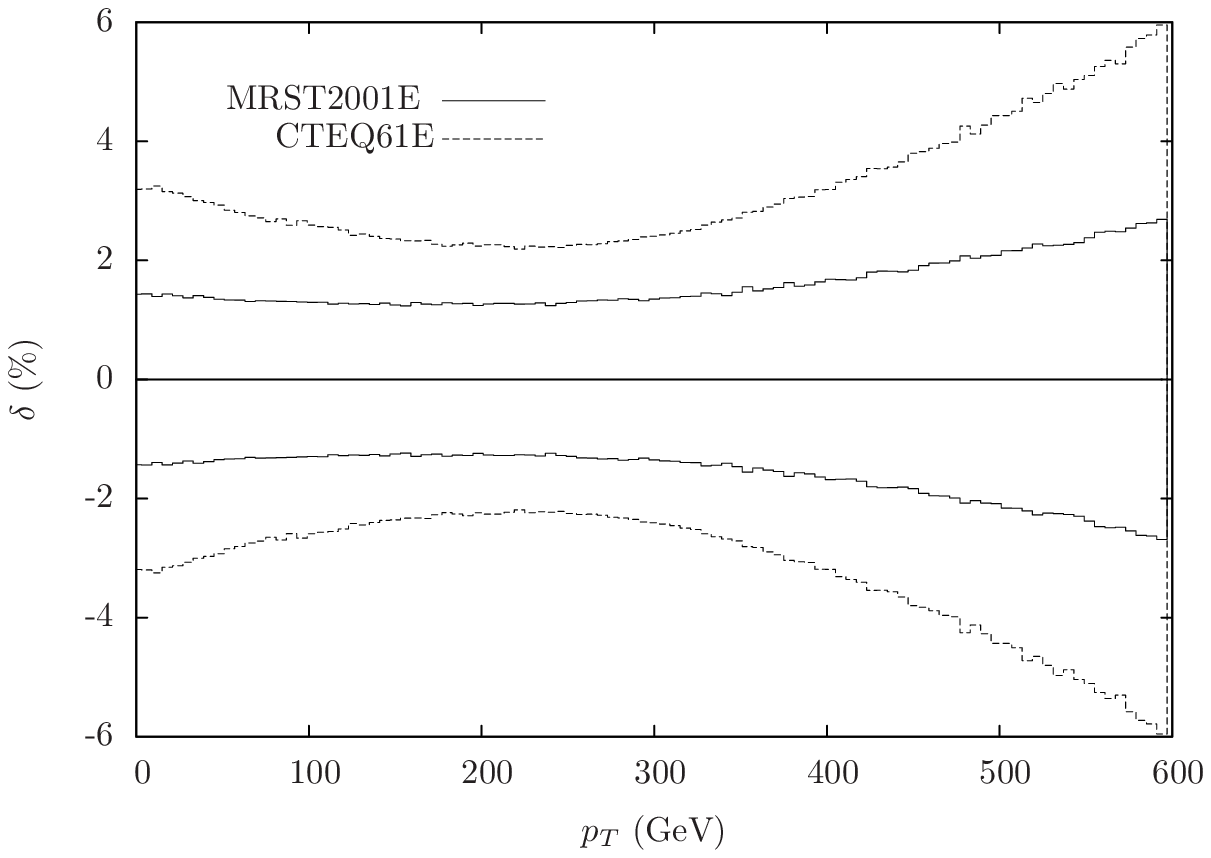 , width=8cm}
\epsfig{file=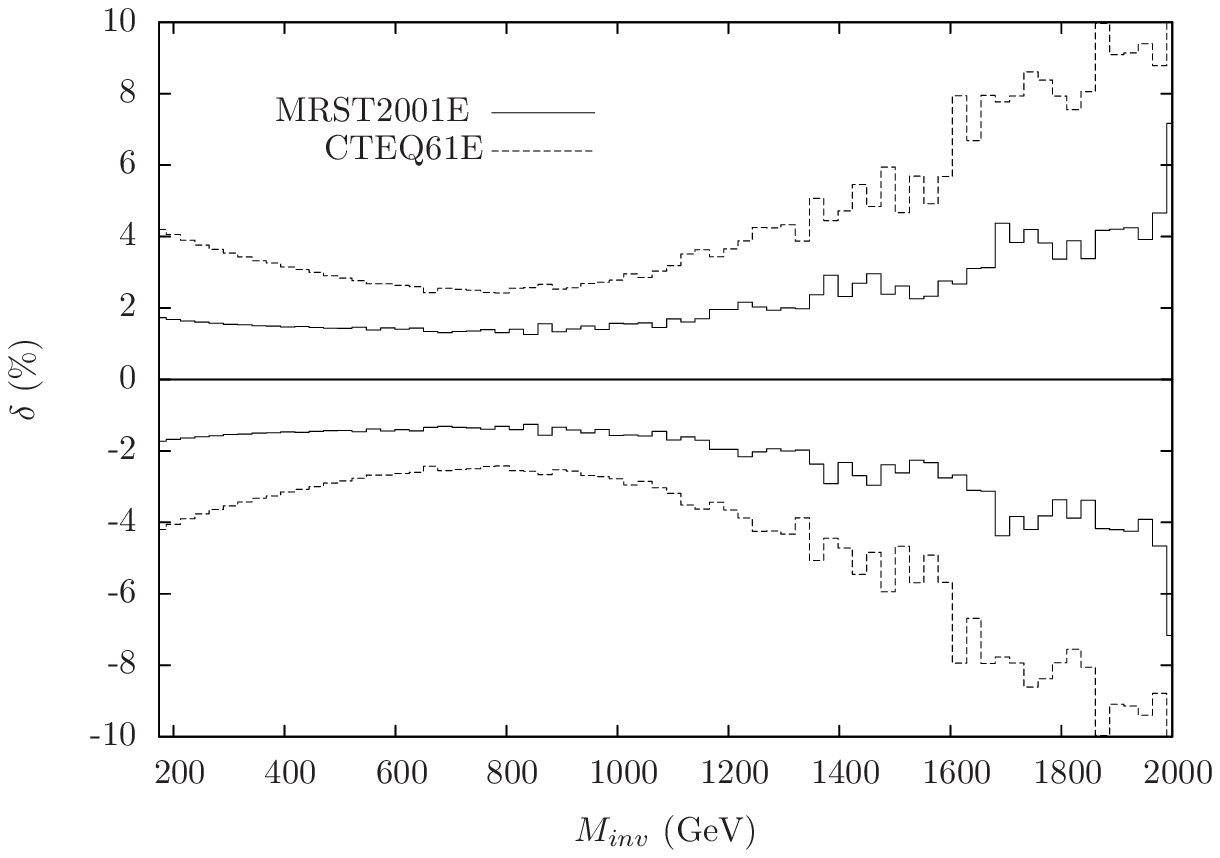, width=8cm}  
\caption{\footnotesize{
Left Panel: 
PDF uncertainty (in percent) on the $\frac{d \sigma}{dp_{T}}$ distribution 
(LO calculation). For each bin, the minimum and maximum deviations 
w.r.t the best fit PDF, as given by the MRST2001E set (solid lines) and 
by the CTEQ61E set (dashed lines), are shown. 
Right: the same as in the left panel, 
for $\frac{d \sigma}{dM_{\mbox{\tiny inv}}}$.}
\label{fig:pdfunc} }
\end{figure}

An important source of theoretical uncertainty is given by the contribution 
of the parametric errors associated with the parton densities. This 
could become particularly relevant for single top channels, due to the 
presence of an initial state $b$ quark, 
whose distribution function is strictly related to the gluon distribution. 
We have studied the impact of such uncertainties on the 
transverse momentum distribution $\frac{d \sigma}{dp_{T}}$ and 
on the invariant mass distribution $\frac{d \sigma}{dM_{\mbox{\tiny inv}}}$ 
by using the PDF sets MRST2001E and CTEQ61E as in the LHAPDF 
package~\cite{lhapdf}. For each bin in the histograms the maximum 
and minimum values are calculated starting 
from the central value according to the formula 
$\sigma({\rm central}) \pm 1/2 \sqrt{\sum_{i=20(15)} 
(\sigma(2i - 1) - \sigma(2i))^2}$, 
according to the prescription of Ref.~\cite{pdfpage}. 
The results are shown in Figure~\ref{fig:pdfunc}.
The spread of the predictions obtained with the MRST set displays a 
relative deviation of about 2\% or less, increasing on the large scale tails 
of the distributions (where the cross section is however very small), 
while the CTEQ set gives a larger 
uncertainty, of the order of 3-4\%. 
This is due to different values of the tolerance parameter~\cite{Tpdf}, the 
latter being defined as the allowed maximum of the $\Delta\chi^2$ variation 
w.r.t. the parameters of the best PDFs fit. 
Conservatively, we can associate to our predictions an uncertainty due 
to the present knowledge of parton densities of about 3\%. It is also worth 
noting that the uncertainties obtained according to such a procedure 
are of purely experimental origin only ({\it i.e.} as due to the systematic 
and statistical errors of the data used in the global fit), leaving aside 
other sources of uncertainty of theoretical origin.

\section{Conclusions} 
\label{sec:concl} 
 
We have computed in this paper the complete one-loop electroweak effect on  
various observables in the MSSM. The calculation has fully included QED 
effects. The overall result is that the one-loop effect is small,  
of a  positive few percent in the total rate that we have considered  
as the first realistically measurable quantity.  
Technically speaking, this small number arises from a 
competition of negative weak contributions and positive QED terms.  
In the considered mSUGRA symmetry breaking scheme, the genuine SUSY  
effect in the considered benchmark points is systematically modest,  
at most of a one percent size. The values that we have obtained e.g. for  
the total rate should still be modified by the additional NLO QCD 
contribution. The latter is known, small and theoretically safe and could  
be easily added to our calculation.  
It will appear in a forthcoming paper that will provide an expression  
for the overall single top production at LHC, including the already  
existing calculations for the associated $tW$ production (our paper and the QCD 
ones).  
We have also given an estimate of the parametric errors associated with 
the present knowledge of the parton densities. The distributions studied 
in this paper are affected by a few percent PDF uncertainty. It is 
worth saying that this uncertainty is expected to be lowered once 
the LHC data will become available. 
In conclusion, a precise measurement of the $t$-channel rate appears as a  
perfect way to determine the value of the $V_{tb}$ coupling,  
both within the SM  
and within the MSSM with mSUGRA symmetry breaking. We cannot exclude, though,  
that for different symmetry breaking mechanisms the genuine SUSY effect is   
more sizable. This question, that is beyond the purposes of this preliminary 
paper, remains open and, in our opinion,  
would deserve a special dedicated rigorous analysis.

\vspace{1cm} 
\noindent 
{\bf Acknowledgements} \\ 
We are gratefull to T.~Hahn an S.~Pozzorini for helpful discussions. E.M. is indebted with S.~Dittmaier for  
valuable suggestions on the Dipole Subtraction Formalism.


 
\newpage 
 
\begin{figure}[htb] 
\centering 
\epsfig{file=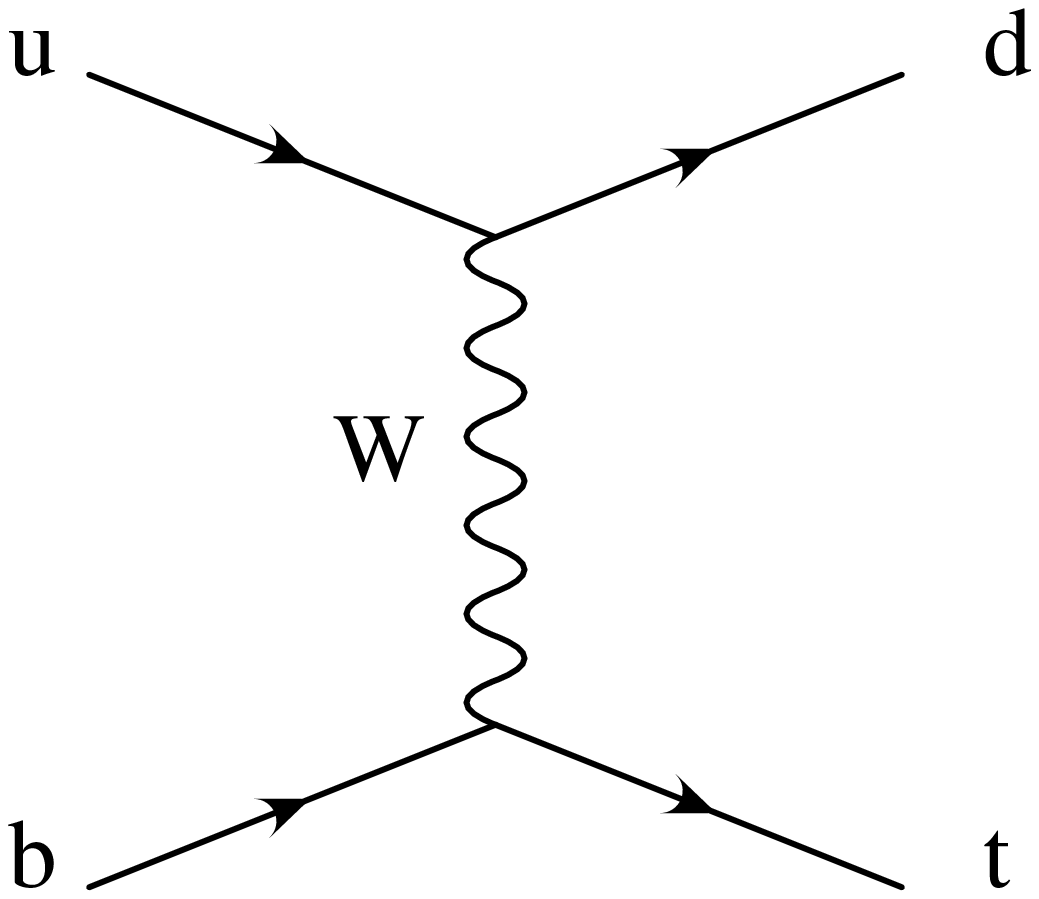, width=6cm, angle=0} 
\ \ \ \ \ \ \ \ \ \ \ \ \ \ \ \  
\epsfig{file=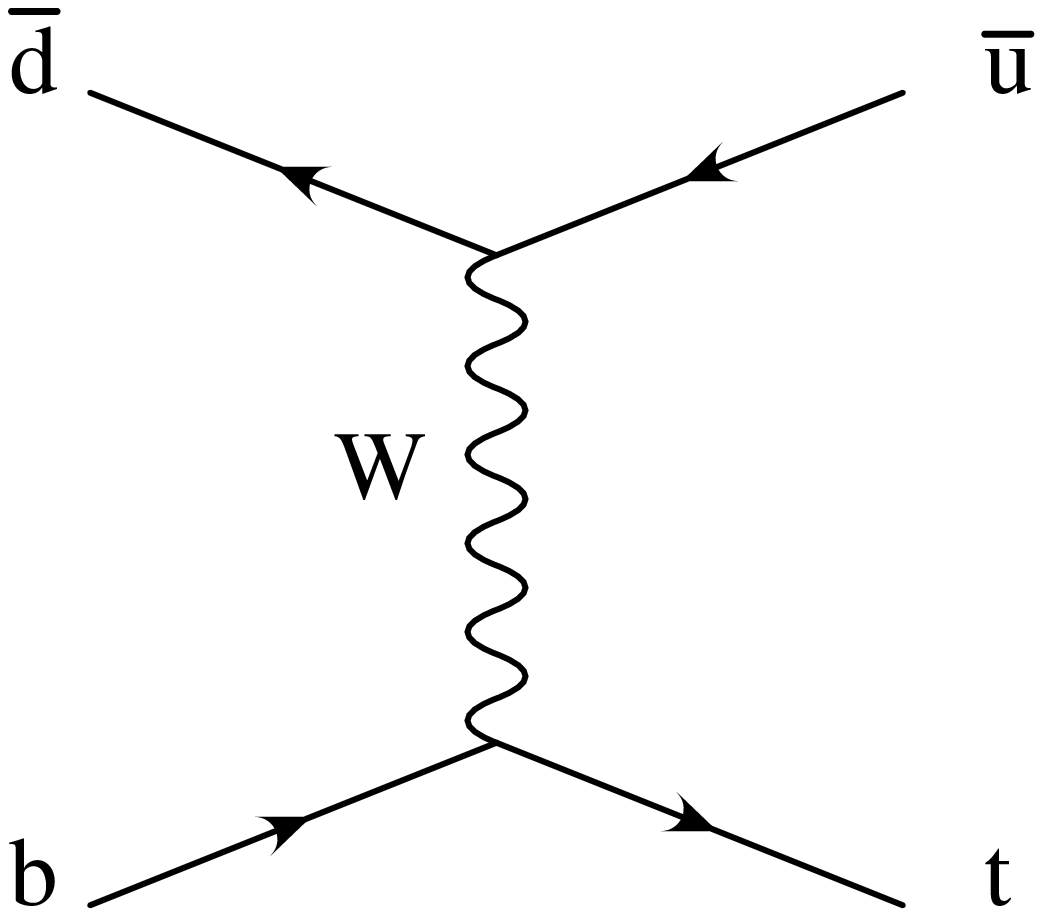, width=6.4cm, angle=0} 
\vspace{1.5cm} 
\caption{Born direct and crossed processes for single top production in the $t$-channel 
with first generation light quark current.} 
\label{fig:1} 
\end{figure} 
 
\clearpage 
 
\clearpage 
 
\clearpage 
 
\clearpage 
 
\clearpage 
 
\clearpage 
 
\end{document}